\documentclass[a4paper,fleqn,usenatbib]{mnras}

\usepackage[T1]{fontenc}
\usepackage{ae,aecompl}

\usepackage{longtable}

\usepackage{graphicx}	
\usepackage{amsmath}	
\usepackage{amssymb}	
\usepackage{lscape}

\usepackage{subfigure}

\title[LMT/AzTEC Mon~R2 Dense Core Survey]{Early science with the Large Millimetre Telescope: An LMT/AzTEC 1.1~mm Survey of Dense Cores in the Monoceros~R2 Giant Molecular Cloud}

\author[A. Sokol et al.]{Alyssa~D. Sokol$^{1}$\thanks{E-mail: asokol@umass.edu},
R.~A. Gutermuth$^{1}$,
R. Pokhrel$^{1}$,
A.~I. G\'omez-Ruiz$^{2}$,
\newauthor
G.~W. Wilson$^{1}$,
S.~S.~R. Offner$^{3}$,
M. Heyer$^{1}$,
 A. Luna$^{2}$,
 F.~ P. ~Schloerb$^{1}$,
 \newauthor
 and D. S\'anchez $^{2}$
\\
$^{1}$Department of Astronomy, University of Massachusetts, Amherst, MA 01003, USA \\
$^{2}$Instituto Nacional de Astrof\'isica Optica y Electr\'onica Luis Enrique Erro \#1, CP 72840, Tonantzintla, Puebla, M\'exico \\
$^{3}$Department of Astronomy, The University of Texas at Austin, Austin, TX 78712, USA
}

\pubyear{2017}

\begin{document}
\label{firstpage}
\pagerange{\pageref{firstpage}--\pageref{lastpage}}
\maketitle

\begin{abstract}
We present a 1.1~mm census of dense cores in the Mon~R2 Giant Molecular Cloud with the AzTEC instrument on the Large Millimeter Telescope (LMT). We detect 295 cores (209 starless, and 86 with protostars) in a two square degree shallow survey. We also carry out a deep follow-up survey of 9 regions with low to intermediate ($3<A_V<7$) gas column densities and detect 60 new cores in the deeper survey which allows us to derive a completeness limit.
After performing corrections for low signal-to-noise cores, we find a median core mass of $\sim 2.1 \text{M}_{\odot}$ and a median size of $ 0.08$~pc. $46\%$ of the cores (141) have masses exceeding the local Bonor-Ebert mass for cores with T=12K, suggesting that in the absence of supporting non-thermal pressure, these regions are unstable to gravitational collapse.  We present the core mass function (CMF) for various subdivisions of the core sample.  We find that cores with masses $>$10~$M_{\odot}$ are exclusively found in regions with high core number densities and that the CMF of the starless cores has an excess of low-mass cores ($<$5~$M_{\odot}$) compared to the CMF of protostellar cores.  We report a power law correlation of index $1.99 \pm 0.03$ between local core mass density and gas column density (as traced by Herschel) over a wide range of size scales (0.3-5~pc). This power law is consistent with that predicted for  thermal fragmentation of a self-gravitating sheet.
Finally, we find the global core formation efficiency increases with gas column density, reaching $\sim$ 43\% efficiency for gas with $A_{V}\geq 30$.
\end{abstract}



\section{Introduction}


The majority of stars form in clusters \citep{Porras2003,LadaLada2003,Allen2007,Bressert2010}. However, the details of the process that ultimately sets the stellar spatial distribution are still poorly understood. 
Ground-based near-IR molecular cloud surveys have characterized the spatial distributions of young stars statistically \citep[e.g][]{Lada1991, Carpenter2000, Gutermuth2005, Roman-Zuniga2008}, finding stellar over-densities relative to unrelated field stars in both dense and diffuse regions of molecular clouds.  These results suggest a variety of formation environments for low-mass stars.  The wide-field, sensitive, mid-IR (3-24$\mu$m) surveys from \textit{Spitzer} enabled the detection of most young stellar objects (YSOs) by virtue of excess IR emission.  With this high confidence membership criterion, stellar densities in nearby molecular clouds have been mapped in greater detail\citep{Allen2007, Evans2009, Gutermuth2009,Gutermuth2011, Megeath2012}. 

Formulating a predictive theory of star formation requires an improved understanding of the role of prestellar cores.  Dense cores are localized volumes of gas with higher density than the ambient cloud. Such sites provide a reservoir of material from which protostars are formed, but the factors that determine the conversion efficiency from reservoir mass to resultant stellar mass are unclear. Improving our understanding of the formation and evolution of dense cores is necessary to shed light on the mass efficiency of star formation under differing conditions. Observations to date indicate that the dense core mass function (CMF) resembles the stellar initial mass function \citep[IMF;][and references therein]{Alves2007, Offner2014}, suggesting that the mass required for forming an individual star may be predetermined at the core stage. If stellar masses are largely determined by the amount of gas in their parent core, the relationship between the two mass distributions gives insight into the processes that are responsible for star formation \citep{Meyer2000}. 

Due to their low temperature (T$<$12~K) and high density ($n > 10^{4}$~cm$^{-3}$), the ideal wavelength range to observe dense cores is the submillimeter to millimeter. At these long wavelengths, thermal emission from dust at core densities is optically thin, and observed flux density traces total mass of dust in cores. Ground-based observations of mm cores have been very successful within the nearby clouds of Gould's Belt \citep{Motte1998,Harvey03, Kirk2005, Enoch2006, Enoch2007, Enoch2008}. More recently, space-based observations have made strides in surveying dense core populations. With the advent of \textit{Herschel} \citep{Pilbratt2010}, cores can be studied with high-sensitivity multi-band imaging in the far-infrared and submillimeter, allowing the CMF to be probed to lower masses. The Herschel Gould Belt Survey (HGBS)\citep{Andre2010} covered 15 nearby molecular clouds with a variety of star formation environments. As part of the HGBS, core populations have been characterized in clouds such as Taurus \citep{Marsh2014, Marsh2016}, Aquila \citep{Konyves10, Konyves15}, Perseus \citep{Sadavoy2012},  and Orion\citep{Kirk2016}. 


At 830~pc away \citep{racine1968}\footnote{\citet{Dzib2016} published a VLBA parallax distance measurement of 893$\pm$40~pc to Mon~R2, largely consistent with the canonical value adopted here.}, the Mon~R2 giant molecular cloud (GMC) is more distant compared to nearby clouds studied in the Herschel Gould Belt Survey. \citet{Gutermuth2011} reported that the cloud has a clear positive correlation between the spatial distribution of YSOs and local gas column density. 
Though many YSOs have been detected in Mon~R2 ($\sim 1000$), no census of dense cores has been performed to date. \citet{Pokhrel2016} presented a Herschel/SPIRE survey of Mon~R2, mapping a dynamic range of density substructure in the cloud. However, the 36\arcsec\ resolution at 500$\mu$m (corresponding to a physical size of 0.14~pc at 830~pc), makes it difficult to separate beam-diluted dense cores (0.05 $\times$ 0.05~pc) from surrounding diffuse cloud structure. 

Here we present the first ever census of dense cores across the entire Mon~R2 GMC (4$\times$10$^4$~$M_{\odot}$) using the AzTEC 1.1~mm camera on the Large Millimeter Telescope (LMT).  The sensitivity, resolution and speed of LMT/AzTEC enables fast, wide field mapping of the dust emission at scales that are essential to properly separate and characterize dense cores.
Our goal is to characterize the core population and spatial distribution, examine the core mass function (CMF) across a wide mass range and differing environments, and probe the effects of gas morphology (as identified by Herschel) on core clustering and formation efficiency throughout the cloud. In $\S$\ref{sec:observations} we describe the LMT/AzTEC observations and core identification process. Our study utilizes data primarily from an initial, spatially extensive, shallow survey in addition to follow-up deep observations of selected regions. In $\S$\ref{sec:coreprop} we make an approximate flux and size correction for low signal to noise ratio (S/N) cores and show the core mass versus size relation. 
In $\S$\ref{sec:cmf} we report the CMF in both clustered and isolated environments. In $\S$\ref{sec:cfe} we look at the relationship between the core mass distribution and surrounding gas properties on local and global size scales to constrain the dependence of core formation efficiency on gas column density in the cloud. We also compare the clustering properties of cores with stellar clustering studied by \citet{Gutermuth2011} and discuss the similarities in their dependence on local gas structure. Finally, in $\S$\ref{sec:summary} we summarize our results.



\section{Observations}
\label{sec:observations}

\begin{figure}

	\includegraphics[width=\columnwidth]{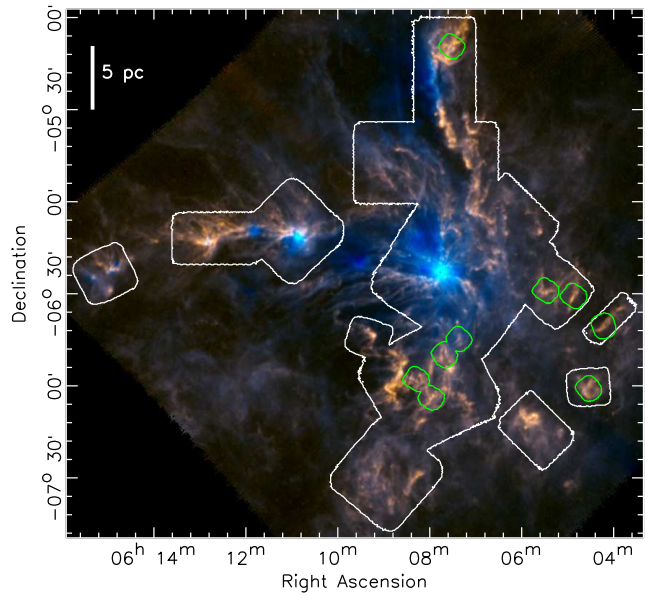}
   \caption{The Herschel SPIRE column-density temperature map of Mon~R2 \citep{Pokhrel2016} with the shallow LMT/AzTEC 1.1~mm survey coverage contours overplotted in white and the deep survey coverage contours in green. Intensity is mapped as column density and colour is mapped as temperature where the redder areas are colder ($<$10~K) and bluer areas are warmer ($>$10~K). }
    \label{fig:coverage}
\end{figure}

\begin{figure*}

	\includegraphics[width=.6\textwidth]{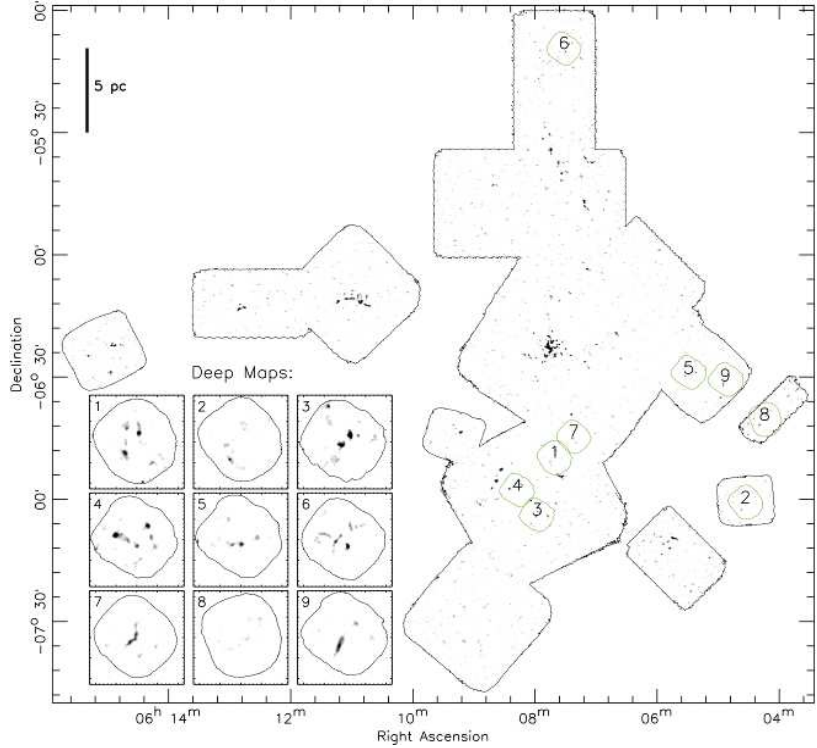}
   \caption{The full set of LMT/AzTEC 1.1~mm survey images. The main figure shows the combined shallow survey signal to noise map in inverse linear greyscale from 2-5 sigma, roughly 14-35 mJy/beam. The inset panels show the nine deep fields with the same scaling style, but ranging from 2-10 sigma, or roughly 6-60~mJy/beam. The locations and indexes of the deep fields are marked on the shallow survey map.}
    \label{fig:aztecdata}
\end{figure*}

We observed Mon~R2 with AzTEC \citep{wilson2008}, the 144-element 1.1~mm bolometer array on the 50-m diameter Large Millimeter Telescope Alfonso Serrano (LMT) in its 32-m diameter early science configuration from 27~Nov.~2014 to 31~Jan.~2015.  Fourteen fields totaling $\sim$2 sq. deg. were chosen based on the Herschel survey of the entire cloud \citep{Pokhrel2016} such that we covered the majority of the cloud area found at $N(H_2)>3\times10^{21}$~cm$^{-2}$ along with adjacent, lower column density areas within our rectangular survey areas.  In addition to the large area shallow survey, we obtained a deeper LMT/AzTEC 1.1~mm survey of selected fields in Mon~R2 from 22~Jan.~2016 to 25~Feb.~2016 with the goal of characterizing the low-mass end of the CMF with higher confidence.  We selected nine regions, each 5\arcmin\ in diameter, that had relatively low mean column density but exhibited filamentary structure in the Herschel map \citep{Pokhrel2016} and had relatively few dense cores detected in our shallow survey.  In Figure~\ref{fig:coverage}, we plot the coverage contour for both surveys over the column density-temperature map derived from the Herschel data \citep{Pokhrel2016}.  The observations are shown in Figure ~\ref{fig:aztecdata}, and the detailed survey information is summarized in Table~\ref{table:obs}.

The queue observing strategy used at LMT ensures that AzTEC programs are generally observed under good 1.1~mm sky conditions ($\tau \sim 0.1$ or better).  For our observations, $\tau$ ranged from 0.02 to 0.11.  The calibration scheme for AzTEC on LMT is similar to AzTEC guest installations on other telescopes \citep[e.g.][]{scott2008}.  In summary, load curves and beam maps were obtained 2-3 times per night and pointing offset calibrations were taken every hour.  The primary flux calibrator and beam-mapping target was CRL618, and the pointing offset calibrator target was quasar 0607-085 for all fields.  Pointing offsets drifted by 1.5-2\arcsec\ per hour.  We interpolate pointing offsets between successive calibrator measurements, thus residual pointing offsets are $<$1\arcsec.

The shallow AzTEC maps were obtained in ``large map mode'', a simple raster scanning scheme whereby the telescope is slewed at a constant scan rate (100\arcsec~s$^{-1}$) in R.A. or Azimuth, and then again over the same area in Dec. or Elevation scan directions to produce roughly perpendicular scan orientations.  The choice of scan coordinate system is set based on the shape of the map.  Azimuth-elevation mapping is used for relatively round or square maps to enhance the variety of scan orientations.  Rectangular maps are scanned in fixed R.A.-Dec. orientation to maximize survey efficiency.  Both mapping schemes achieved similar noise levels per integration time, typically achieving noise levels of $\sim$7~mJy per beam RMS.  Noise levels are determined from the time stream data using a jackknifing technique \citep{scott2008}.  We trimmed the ends of raster scans where the scan speed and direction did not match the main mapping area, potentially inhibiting the atmospheric filtering algorithm.  

In contrast, the deep AzTEC maps were obtained in ``rastajous'' mode, a raster mapping scheme combined with a lissajous scanning pattern.  This mode is specifically designed to improve observing efficiency on $\sim$7\arcmin\ fields, where the LMT's standard lissajous mapping pattern is too small, and raster mapping would result in increased overheads in time owing to reversing the motion/direction of the antenna. The resultant noise level was $\sim$3~mJy per beam for the deep maps.

All the data were processed following the standard AzTEC data treatment, including ``despiking" of the signal time streams, low-pass filtering, and principle component analysis (PCA) atmospheric filtering \citep{scott2008}.  Adaptive Wiener filtering is commonly used on data like these to convolve the filtered data optimally for revealing unresolved sources without sacrificing angular resolution.  The excellent angular resolution of AzTEC on LMT corresponds to a physical resolution of 0.03~pc at the distance to Mon~R2.  However, since typical dense cores are $0.05 \times 0.05$~pc \citep{Enoch2007}, Wiener filtering results in substantial loss of flux in our targets of interest.  Thus, we replaced the Wiener filter with a simple Gaussian filter of the same FWHM as the telescope beam (8.5\arcsec).  The result is a marginal loss of angular resolution in the final images (12\arcsec\ FWHM, 0.05~pc at 830~pc distance) and a substantial improvement in the number and S/N of detected cores.  In addition, it is common usage with our PCA atmospheric filtering scheme for the user to remove 7-9 of the highest power eigenvectors of correlated emission.  Such strong filtering also trims spatial scales of flux variation that can affect measurements of our marginally resolved cores.  Thus we tested a range of lighter PCA filter settings and chose to remove only the first three eigenvectors, yielding the greatest total signal to noise in our test fields.  


\subsection{Core Identification}
\label{sec:coreid}

The primary goal of this survey is to provide the first ever census of dense gas cores in the Mon~R2 GMC.  We adopt a core identification process that is similar to the classic two dimensional Clumpfind algorithm \citep{williams1994}.  Specifically, we first mask the flux map to allow only those values with $S/N > 2.5$ and integrated coverage greater than 40\% of the median non-zero coverage value of each map.  Then we scale the masked signal map by the median noise value in the masked pixels and truncate fractional values to yield a noise-based contour image, with steps of $\sim$1~$\sigma$ in size.  That map is inverted and fed to the native IDL function {\it watershed} to identify the footprints of contiguous zones of emission in the map.  Any substantial saddle points in otherwise contiguous regions of emission are used to divide them into separate objects with single peaks.  The resulting contiguous region footprints mark the boundaries of the candidate dense gas cores.  Total footprint area, footprint centre position (not flux-weighted), total flux in the footprint (and S/N), half peak power area, half peak power centre position (flux-weighted), and peak flux per beam (and S/N) are measured for all candidates.  Finally, we reject all core candidates whose footprints fall within 8\arcsec\ of the coverage edge. This step eliminates a major source of false detections and excludes cores that may not have been fully covered within the survey area.

\begin{figure}
	\includegraphics[width=\columnwidth]{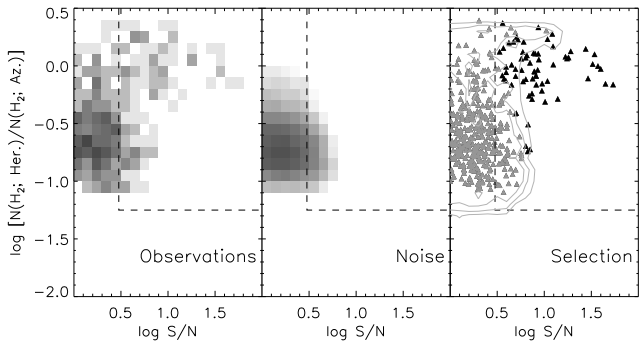}
   \caption{Column density ratios (Herschel over AzTEC) versus Total S/N for core candidate selection, using Field~11 as an example.  The left and centre panels are 2D histograms for the core candidates from observations and the noise realizations, respectively.  The dashed lines mark our fixed limits at S/N$> 3$ and log ratio $> -1.25$, the maximal beam dilution value.  The right panel shows the observed candidates and a contour mapping of the ``goodness'' score, the ratio of the observed to the noise-derived 2D histograms, at levels of 0.5, 0.75, and 0.9.  Points in black fall above the 0.75 contour, our requirement for the final core selection. }
    \label{fig:core_select1}
\end{figure}

\begin{figure}
	\includegraphics[width=\columnwidth]{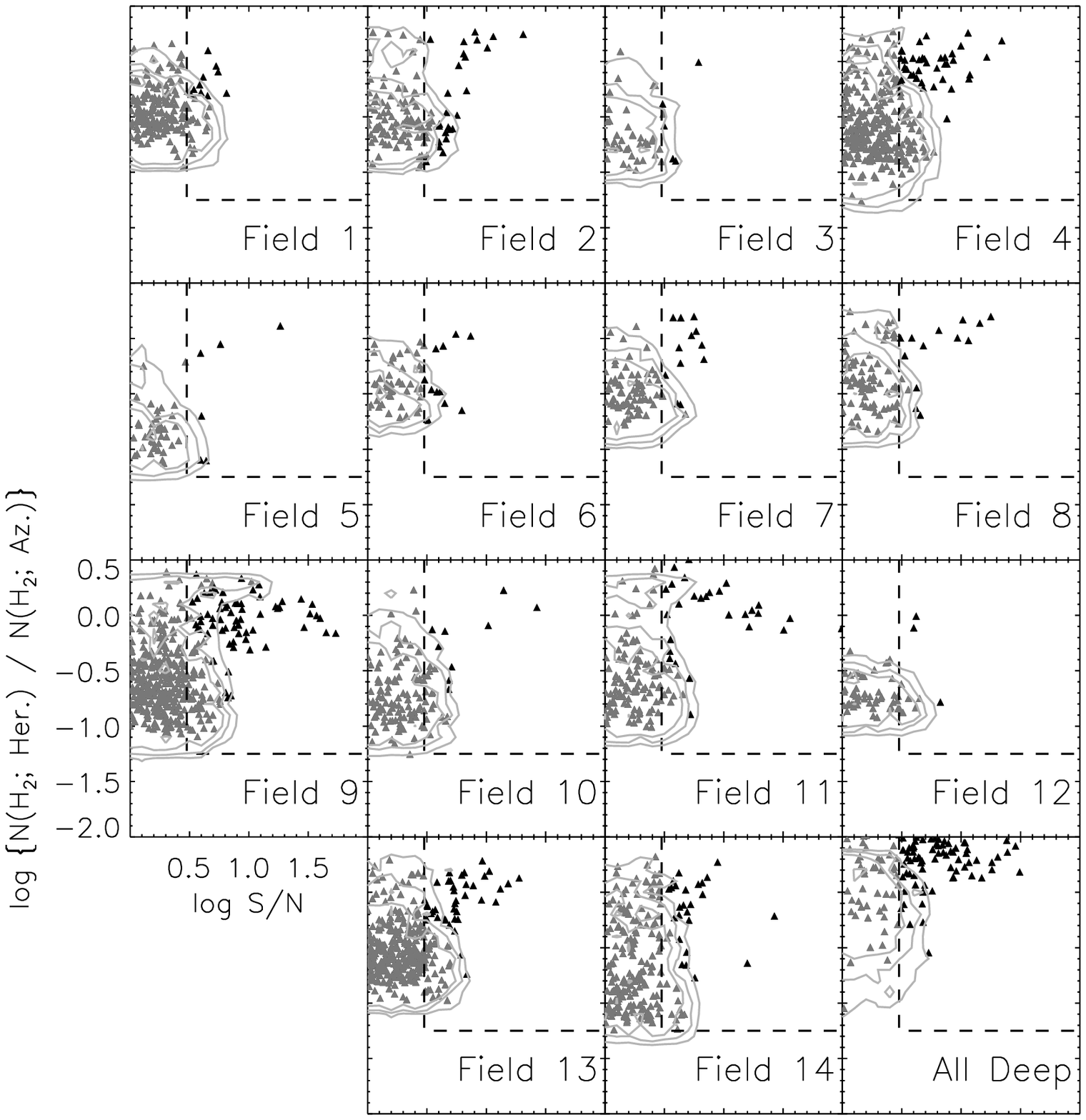}
    \caption{Column density ratios (Herschel over AzTEC) versus Total S/N for core candidate selection in all fields, along with the deep mapping combined in the last panel. Overlays match the right panel of Fig.~\ref{fig:core_select1}}
    \label{fig:core_select1b}
\end{figure}

A jackknifing process is applied to the time-stream data to eliminate any astronomical signal and produce an empirical noise characterization.  This process yields fully filtered and spatially mapped noise realizations (100 per field) that are generated identically to the observations.  We passed these realizations through our core identification process to estimate the false detection rate for a given choice of core candidate detection requirements.  For all cores, observed candidates and noise-realized false cores, we estimated their mean dust column density from the AzTEC emission and compared it to the Herschel-derived dust column density at the same position.  As shown in Figure~\ref{fig:core_select1}, most high S/N cores observed in this survey occupy a narrow range in their ratio of the AzTEC- and Herschel--derived mean column densities.  In contrast, false cores from the noise realizations correlate poorly with the diffuse gas distribution.  We construct 2D histograms for both the observed core candidates and those detected in the noise realizations.  For every observed core candidate, we assign a confidence score based on the ratio of the 2D histogram bins (see Table~\ref{table:allcores}) observed over normalized noise at that position.  All observed cores with less than a 25\% chance of being consistent with a false detection are used for our final core selection, yielding a total of 295 cores selected in the shallow survey (see Fig.~\ref{fig:core_select1b} for the selection diagrams of all fields). Integrating over the contamination probabilities of those objects, we estimate that 21 of those are residual contaminants, a false detection rate of $\sim$7\%. Under the same identification processes, the deep survey yields a final list of 79 cores, with an estimate of 1 residual contaminant core after integrating over contamination probabilities ($\sim 1.3 \%$).  In Appendix~\ref{sec:appendixtests} we test our false detection rate in a comparison between core detections and noise levels in cloud regions with overlapping deep and shallow maps. 

In this paper, we refer to any core lacking a IR-excess YSO projected on its AzTEC footprint as \textit{starless}. Of the 295 cores in the shallow survey, we identify 209 as starless and 86 with stars (71\% starless fraction). Additionally, 65 of the 79 cores identified in the deep survey are starless (82\% starless fraction).  
Cores with an internal luminosity source have both 1.1~mm emission and far-infrared continuum emission that has been absorbed and reradiated by a cold dusty envelope. Thus, protostars with substantial envelopes, commonly referred to as Class 0 phase YSOs \citep{Andre1993}, should largely fall in this category.  In our survey, cores with protostars are sources with an overlapping Spitzer-identified YSO. This can result in some chance super-positions of YSOs with cores, but the small angular size of our cores should make this a relatively rare occurrence. 

\begin{figure*}
	\centering
	\includegraphics[width=\textwidth]{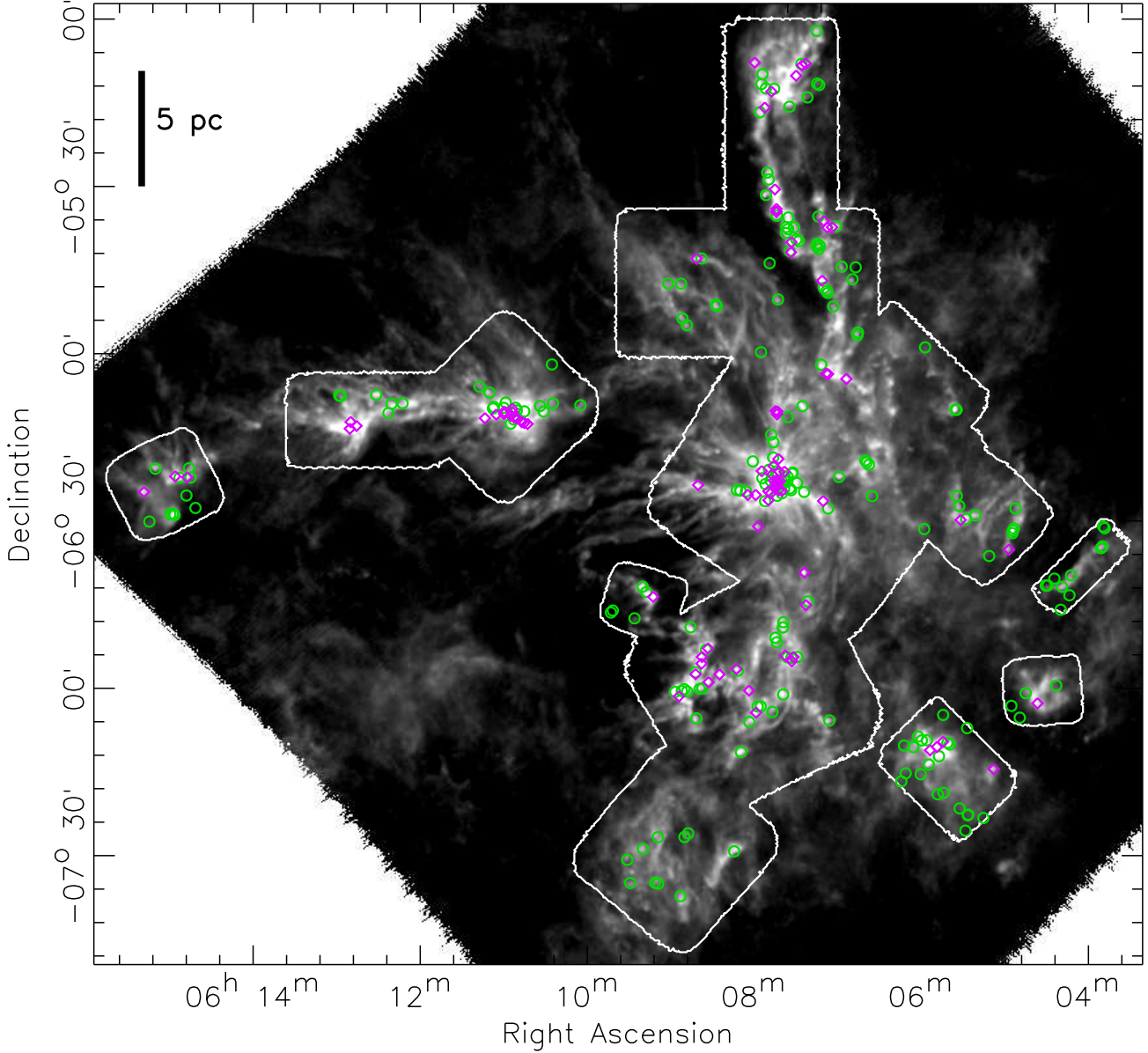}
   \caption{The Herschel greyscale column density map from Figure \ref{fig:coverage} with the core spatial distribution from the shallow survey overplotted. Of the 295 cores detected, we identify 209 as starless, shown as green circles, and 86 with protostars, shown as pink diamonds. Protostellar cores have an overlapping Spitzer-identified source, indicating excess IR emission from a central protostar. The white contours indicate the regions of Mon~R2 covered in our shallow LMT/AzTEC survey. }
    \label{fig:covcores}
\end{figure*}

The spatial distribution of cores detected in the shallow survey can be seen in Figure~\ref{fig:covcores} overplotted on the Herschel-SPIRE column density map \citep{Pokhrel2016}. The contour represents the AzTEC shallow survey coverage. Overall, starless cores and cores with protostars are well-mixed. The core distribution follows the filamentary structures that extend away from the dense Mon~R2 centre, with additional isolated cores lying in the outskirts along less prominent filamentary structure. The active cloud centre contains a high concentration of cores, both protostellar and starless, and is also one of the highest temperature regions in the cloud (see column density-temperature map in Figure \ref{fig:coverage}). 

\subsection{Completeness Decay Characterization}

\begin{figure}
	\includegraphics[width=\columnwidth]{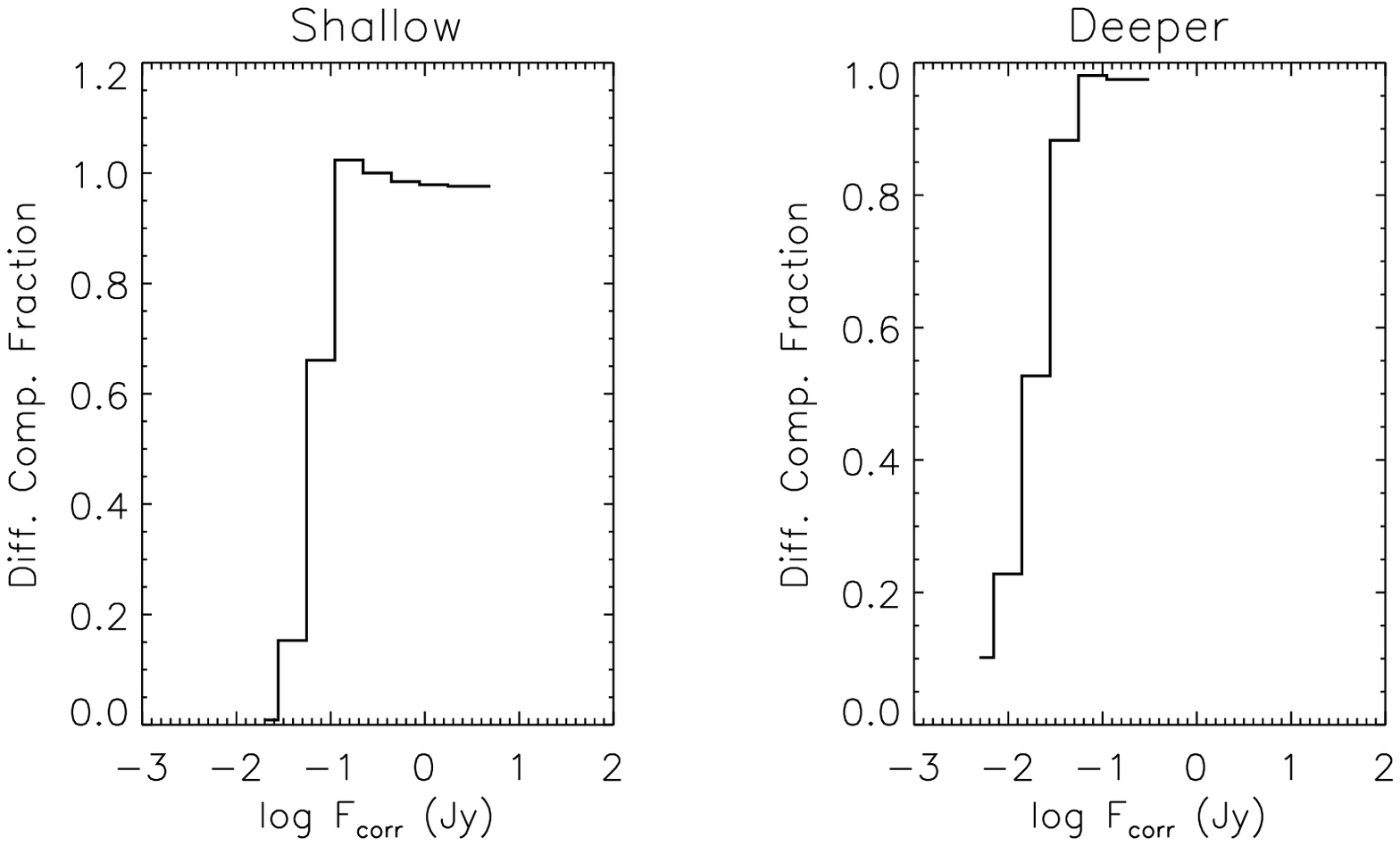}
   \caption{The overall differential completeness fraction as a function of corrected core flux for the shallow and deep surveys, respectively.  This characterization combines the net effects of the partially counteracting biases of traditional detection incompleteness based on artificial core tests and estimates of artificially enhanced completeness due to the inferred presence of some core false detections based on the noise realizations and the core selection process above.}
    \label{fig:corecomp}
\end{figure}

We present differential core detection completeness as a function of corrected core flux for the two surveys in Fig~\ref{fig:corecomp} (Flux corrections are described in $\S$\ref{sec:corepropFLUX}, below).  These measurements are made using a loose grid of nine false cores per mass bin added to the time-stream data for all fields that are then re-reduced using the full filtering pipeline.  Artificial cores are equivalently added to the Herschel column density map so that the identical core extraction and selection process can be applied to the completeness simulation data without loss of generality.  In addition to simple detection completeness, we divide the raw completeness trends by the mean of the ``good'' probabilities for cores within each mass bin to account for false core contamination as a source of artificially enhanced completeness. In summary, the 50\% differential completeness limits (in both corrected flux and core mass) for the shallow and deep surveys are 63~mJy (1.1~$M_{\odot}$) and 18~mJy (0.33~$M_{\odot}$), respectively. 

\section{Core Properties}
\label{sec:coreprop} 

In this section we report the mass and size properties of identified cores, making approximate corrections due to observed systematic effects of a shallow survey. To assess the effectiveness of our corrections, in Appendix~\ref{sec:appendixtests} we perform tests comparing uncorrected and corrected properties of cores detected in both deep and shallow surveys.


\subsection{Flux correction}
\label{sec:corepropFLUX}

\begin{figure}
	\includegraphics[width=\columnwidth]{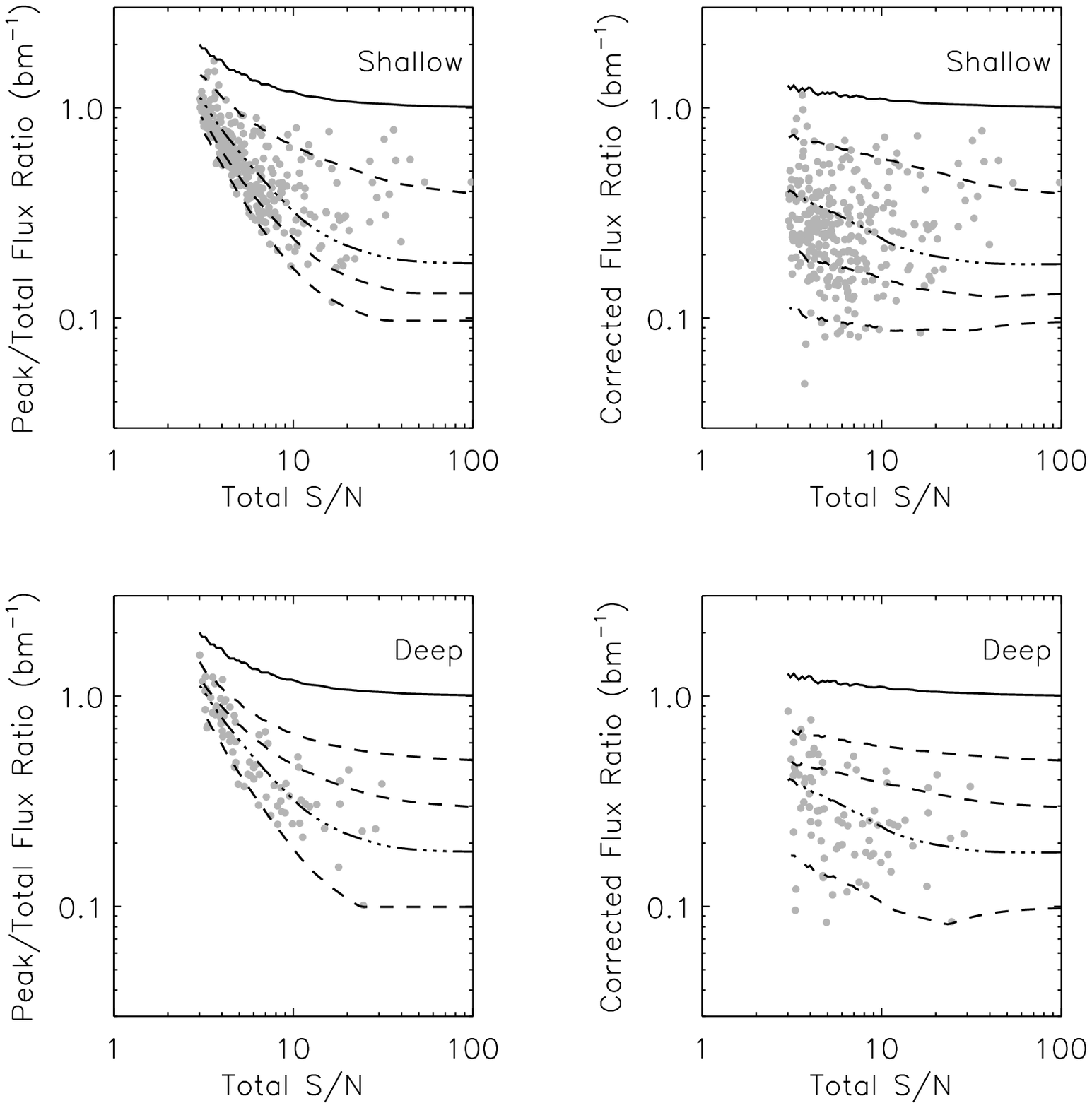}
	\caption{Peak to Total Flux Ratio versus Total S/N for all selected cores. The right column shows the observed flux ratios corrected for noise floor bias, while the left column shows the uncorrected flux ratios. In all plots, the final beam profile is overlaid as a solid line, and the composite core profile is overlaid as a dot-dashed line.  The top row shows the shallow survey core data, with Plummer-like core models of aspect ratio of 2, power law index of 2, and scale lengths of 2, 12, and 18\arcsec\ (from top to bottom) overplotted as dashed lines.  The bottom row shows the deep survey core data, overlaid with Plummer-like core models of aspect ratio of 2, scale lengths of 7.5\arcsec\, and power law indexes of 1, 3, and 5 (from bottom to top) as dashed lines. }
    \label{fig:peaktotal}
\end{figure}

The peak-to-total flux ratios as a function of total signal-to-noise (S/N) are shown in Figure~\ref{fig:peaktotal}. The ratio declines sharply and levels off as S/N increases. It is likely that this fall-off is an effect of a shallow survey of spatially resolved sources, such that lower S/N cores are not detected out to their full radial extents. High ratios at low S/N are the result of this ``noise floor bias'', and are indicative of an underestimation of total flux and consequently an underestimation of core mass. However, the spread in peak-to-total flux ratios for well-detected cores suggests that these ratio values should not be constant at high S/N but are rather an effect of a core's radial profile shape and central concentration. For example, a well-detected core with a higher peak-to-total flux ratio could simply be less extended than its well-detected counterpart with a lower ratio. Such a core could have a more peaked radial profile.   We attempt to correct for this systematic effect by characterizing the observed relationship with an appropriate model and using this as a basis for our total flux correction.

We adopt a simple model for peak-to-total flux ratio as a function of total signal-to-noise ratio,

\begin{equation}
	R(S/N) = F_{peak}/F_{tot} = \delta_R + F_{peak}/F_{corr}
	\label{eqn:ratiomodel}
\end{equation}

\begin{equation}
	\delta_R =  5.25 \times (S/N)^{-1.8}
	\label{eqn:deltaR}
\end{equation} where $S/N$ is the total signal-to-noise ratio of the core within the footprint, $R$ is the observed peak-to-total flux ratio at a given $S/N$, and $F_{peak}/F_{corr}$ is a constant, the ideal value of the peak to total flux ratio at high S/N that varies as a function of the underlying core shape, as discussed above.  With this model in place, we can rearrange Eqn.~\ref{eqn:ratiomodel} and compute the corrected total flux as 

\begin{equation}
F_{corr} = F_{peak} \bigg[ \frac{F_{peak}}{F_{tot}} - \delta_R\bigg]^{-1}
\end{equation} where $F_{corr}$ is the corrected total flux, $F_{peak}$ is the observed peak flux, $F_{tot}$ is the observed total flux, and $\delta_R$ is the S/N-dependent bias term of R(S/N) in Eqn.~\ref{eqn:deltaR}.  

To gauge the performance of the correction for a range of core shapes, we construct several model core radial profiles.  A Plummer-like model \citep{plummer1911} is commonly used to fit pre-stellar cores \citep{Harvey03,VanLoo2014}; it has a flat central profile that steepens to a power-law at larger radii.  Specifically, our Plummer model cores have radial profiles described by:

\begin{equation}
	r^2_{stretch} = ( \rm r \rm cos \theta)^2 \times (a/b) + (\rm r \rm sin \theta)^2 / (a/b)
\end{equation}

\begin{equation}
	\Sigma(r) = \Sigma_0 \times (1+r^2_{stretch}/r^2_{scl})^{-p/2}  \\   
\end{equation} where $\Sigma_0$ is the projected peak surface density, $r_{scl}$ is the scale length, $p$ is the power law index, and $a/b$ is the projected aspect ratio, which we fix at a value of 2 for all models.  We then adopt a range of parameters that subtend the range of the observed core S/N and total flux values.
 
The Plummer-like models, the final beam shape, and all of the actual core observations were passed through this correction model.  The results before and after correction are plotted in Figure~\ref{fig:peaktotal}. In summary, noise floor bias can reduce the total flux of a resolved source by up to an order of magnitude, and the adopted correction reduces this bias to within a factor of 2 of the optimal value for a reasonable range of radial profile shapes, including the beam itself.  

At millimeter wavelengths, the thermal dust emission becomes optically thin, and total flux density traces the total core mass. We derive core mass from the flux assuming modified blackbody emission with a temperature T$=$12~K and a gas to dust mass ratio of 100. We adopt dust opacity value, $\kappa_{\nu}$=0.0121~cm$^2$~/~g, from \citet{OH94} dust model four at $\nu=1100 \mu$m to yield a conversion factor of 18 from Jy to solar masses.
 \subsection{Size correction}
\label{sec:corepropAREA}
 
\begin{figure}
	\includegraphics[width=\columnwidth]{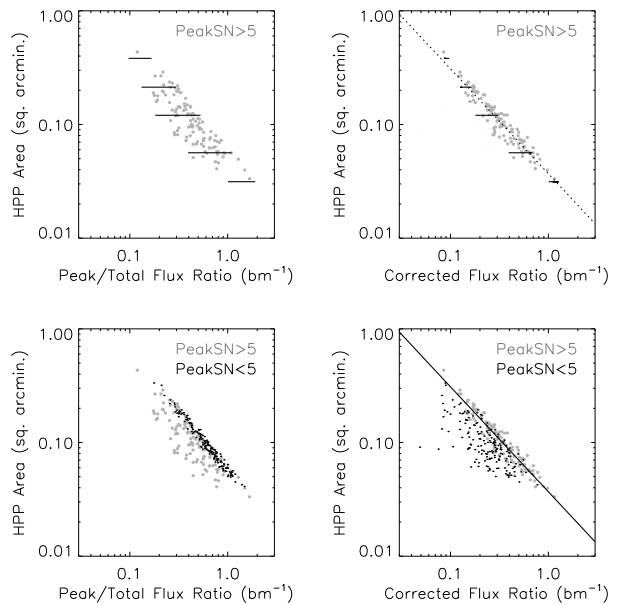}
	\caption{Half-peak power area versus peak to total flux ratio.  The left and right columns use the measured and corrected flux ratios, respectively. The top row of plots show only the reference sample, those cores with a peak S/N $> 5$, in grey points.  These are overlaid with a set of reference models as solid horizontal lines.  The bottom row of plots add the cores with peak S/N $<5$, those that need HPP Area correction, as black points.  The right side plots show the best fit line to the bright subset as a dotted line in the top right and a solid black line in the bottom right plot.  This line is used to determine the HPP Area for the dim cores based on their corrected peak to total flux ratio.}
    \label{fig:hpratios}
\end{figure}

We follow the example of \citet{Konyves15} in reporting our core widths as a deconvolved FWHM size defined as $\text{Size}_{\text{deconv}} = \sqrt{FWHM^{2} - HPBW^{2}}$, where the FWHM is the diameter calculated from the Half-Peak Power (HPP) Area and the HPBW is the 12\arcsec\ AzTEC beam width. 
Since our core search criterion included a mask at S/N$>2.5$, any core with a peak S/N $<5$ will have a HPP Area biased to smaller areas.  
Figure~\ref{fig:hpratios} shows the HPP areas of cores with peak S/N $>5$ descending as a function of peak to total flux ratio, and a tight locus is apparent when the corrected ratio as a function of S/N is plotted instead.  We fit the locus of points to correlate the corrected peak-to-total flux ratio to the measured HPP area in cores with peak S/N $>5$.  Then we use the corrected $F_{peak}/F_{tot}$ of the dim cores (i.e. with peak S/N $<5$) and the fit result to compute the unbiased HPP Area, $A_{HPP,corr}$, 
$$
\text{log} (A_{HPP,corr})=-0.9228 \times log\bigg(\frac{F_{peak}}{F_{tot}}-\delta_R \bigg) - 1.432
$$
where $F_{peak}/F_{tot}$ is the observed, uncorrected peak-to-total flux ratio, $\delta_R$ is the peak-to-total ratio correction term from Eqn.~\ref{eqn:deltaR}, and $A_{HPP,corr}$ is in units of sq. arcmin. As to not further diminish a low S/N core's area, the final HPP area correction is applied to cores with peak S/N <5 \textit{and} $(A_{HPP,corr}- A_{HPP,obs})/A_{HPP,obs} > 0.5$. We utilize cores dually detected in the overlapping deep survey regions to further assess the total flux and size corrections in Appendix~\ref{sec:appendixtests}. 

\subsection{Mass vs. Size}

\begin{figure*}
	\includegraphics[scale=.4]{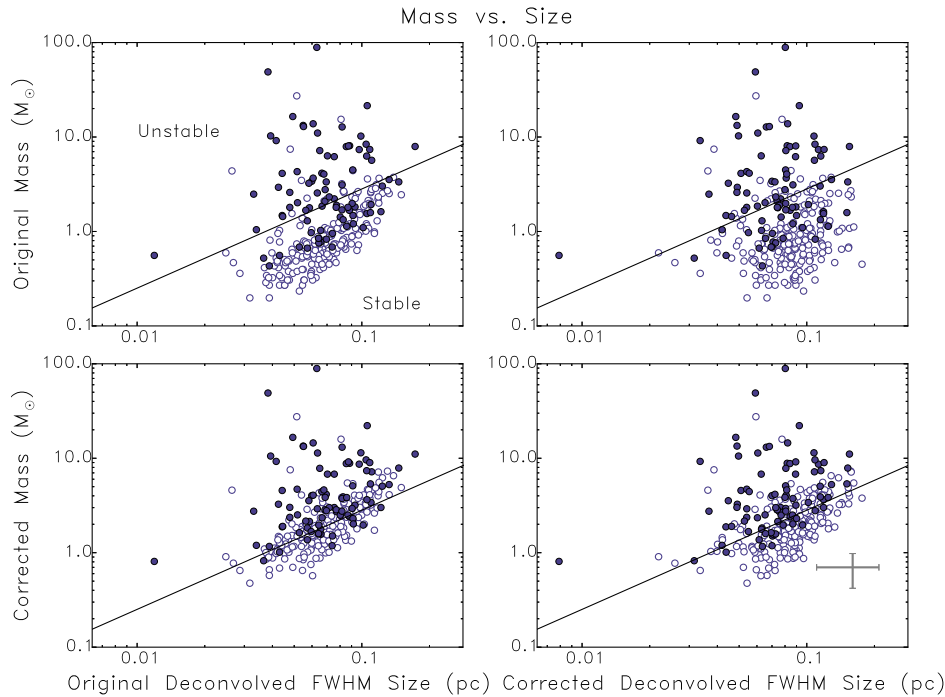}
    \caption{The mass vs. deconvolved FWHM size relation for 295 cores in Mon~R2 at each stage of our flux and size correction, where masses have been derived from total flux (see text for details). In each plot the black line shows the numerically modeled Bonnor-Ebert stability line for cores with T=12K \citep{Konyves15}. Filled in circles represent cores with protostars and unfilled represent starless cores. The upper left plot shows the observed correlation without corrections. The corrected low S/N cores move to higher mass and larger deconvolved FWHM size in the bottom right plot (intermediate stages in the upper right and lower left plot). 56\% of cores are below the Bonnor-Ebert stability line after approximate corrections, most of which (90\%) are starless. The majority of the cores are consistent with being on or above the Bonor-Ebert line to within one sigma uncertainty, implying stability in the form of either gravitational or pressure confinement.}
    \label{fig:massvsize}
\end{figure*}

The mass versus size relation is useful to characterize the self-gravitating stability of prestellar cores. We plot the mass versus size relationship for all 295 identified cores in Figure~\ref{fig:massvsize} at intermediate stages of corrections for low S/N cores. We see the outlying small and low-mass cores (upper left plot) shift upwards and over in the final corrected plot (bottom right). The median uncorrected size is 0.068~pc, while the median corrected size is 0.082~pc.  The median uncorrected core mass is 1.0 $M_{\odot}$ and the median corrected mass is 2.1 $M_{\odot}$.  The black line indicates the approximate numerically modeled Bonnor-Ebert (BE) stability criterion for cores with $T=12K$ \citep{Konyves15}; for higher dust temperatures the BE mass would be larger for a given radius, shifting the mass dependence upward uniformly in log space.  It has been found that a Bonnor-Ebert sphere \citep{bonnor, ebert}, i.e. an isothermal gas sphere in hydrostatic equilibrium, can successfully approximate prestellar cores \citep{all01}.  

We find that of 295 cores, 131 (44\%) are above the stability line, suggesting they are unstable to further collapse. Of the 164 cores below the line, 148 (90\%) of them are starless and 16 have protostars. Surface and cloud pressure are thought to play a significant role in true core stability \citep{Kirk2017}, so it is common to find protostellar cores that may be unbound. However, within uncertainties the majority of cores are consistent with the Bonor-Ebert line, implying that these sources are bound to an extent and stable against further collapse. In this study it remains unclear whether the source of confinement is pressure or gravity; additional velocity information is needed to confirm this approximation and determine the true nature of stability for these cores.

\section{Core Mass Function}
\label{sec:cmf}

The mass distribution of cores formed within a star-forming cloud is theorized to have some bearing on the resultant stellar IMF \citep[e.g.][]{Alves2007, Offner2014}. However, identifying a parallel between core and stellar mass functions is challenging and largely dependent on the nature of sources included.  For example, a CMF comprised of only bound sources may vary from a CMF built with only starless cores. Additionally,  core properties that determine the mass of the resultant star or stars are poorly constrained yet remain crucial to understanding  the link, if any, between the CMF and IMF. Establishing the shape of the CMF has been fraught as well, as systematic effects such as environmental differences and observational biases are often difficult to characterize in most core surveys.  Given the number of cores reported in this work, we can explore some systematic effects for the Mon~R2 CMF caused by these biases, as their differences do affect the resultant CMF shape.

\begin{figure}
	\includegraphics[width=\columnwidth]{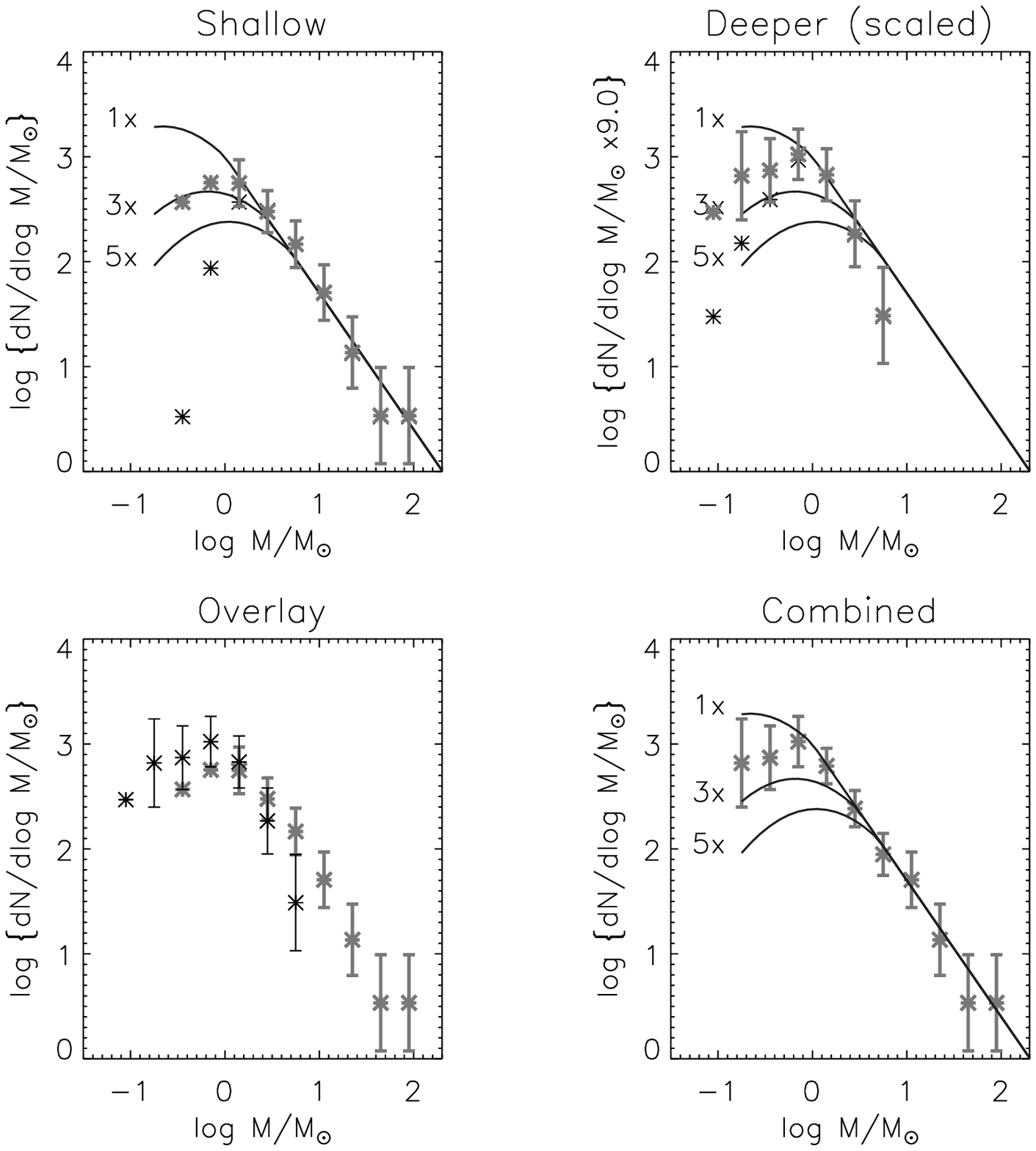}
    \caption{CMFs for all cores in the shallow and deep survey catalogs, and the combined CMF which spans three orders of magnitude in mass.  The top row of plots presents the survey CMFs before and after completeness correction in black and grey points, respectively.  Error bars are plotted only for those points in the corrected CMFs that have $>$20\% completeness. The bottom row demonstrates the shallow and shifted deep CMFs overlay (black and grey points are deep and shallow, respectively) and the merged CMF.  \citet{Chabrier2003} IMFs, with mass scale shifted by factors of 1, 3, \& 5 from top to bottom and power law portions aligned, are overplotted on all but the overlay panel for reference.}
    \label{fig:rawcmfs}
\end{figure}

We present the raw CMFs from the shallow and deep surveys separately in the top row of panels in Fig.~\ref{fig:rawcmfs}.  The shallow survey is relatively unbiased spatially, but its limited depth stifles completeness below $\sim$1~M$_{\odot}$, where one would expect a CMF turnover from the high mass power law to be located, assuming the CMF and IMF shapes are similar.  In contrast, the deeper survey has a strong spatial bias toward subregions of low to intermediate column density (3-7~$A_V$) and sufficient depth to be largely complete in those areas to $\sim$0.3~M$_\odot$.  As explored further in $\S$\ref{sec:cfe}, the core masses in Mon~R2 are systematically higher in denser environments, thus an additional difference in the deeper data is the explicit lack of any cores with masses greater than $\sim$4$M_{\odot}$.  The completeness decay trends in $\S$\ref{sec:observations} are then divided into the CMF of each respective survey to correct for systematic bias as a function of core mass.  Only bins with final completeness greater than 20\% are included in the corrected CMF.  

There is sufficient overlap between the corrected shallow and deep CMFs in the mass range of 0-0.5~dex to compute a scale factor (9.0 $\pm$ 1.8) to make the two consistent over that range.  The scale factor and its Poisson uncertainty are derived from the ratio of the amount of cores in each survey within this mass range, corrected by the relative fractional completeness values. This scaling difference accounts for the much smaller area and typical core clustering environment of the deep survey relative to the respective properties of the shallow survey. The bottom row of panels in Fig.~\ref{fig:rawcmfs} demonstrate a combined CMF.  We adopt the unique bins of each CMF and compute noise-weighted mean values for the overlapping bins.  The result is a combined CMF measurement that extends the viable mass range of each survey by an order of magnitude.  Unfortunately, the combined CMF has implicit blindness to local environmental differences, e.g. local core clustering and ambient column density, for sub-solar mass cores because of the fairly uniform environments of the deep survey fields.  Thus, for analysis of CMF differences with respect to core clustering, we use the CMF derived from the spatially complete shallow survey data only.  

\subsection{CMF Analysis}

The combined CMF for all cores in Mon~R2 in Fig.~\ref{fig:rawcmfs} is consistent with a \citet{Salpeter1955} IMF power law that turns over near 1-3~$M_{\odot}$, e.g. a \citet{Chabrier2003} system IMF with a mass shift by a multiple of 1-3.  The classic shape reported in the CMF literature is Chabrier with a mass shift of 3 \citep[e.g.][]{Alves2007,Andre2010}.  Starting from this basic level of agreement with previous results, we perform some limited exploration of CMF variations as a function of local environment which is possible because of the substantial number of cores found in our surveys.

\begin{figure}
	\includegraphics[width=\columnwidth]{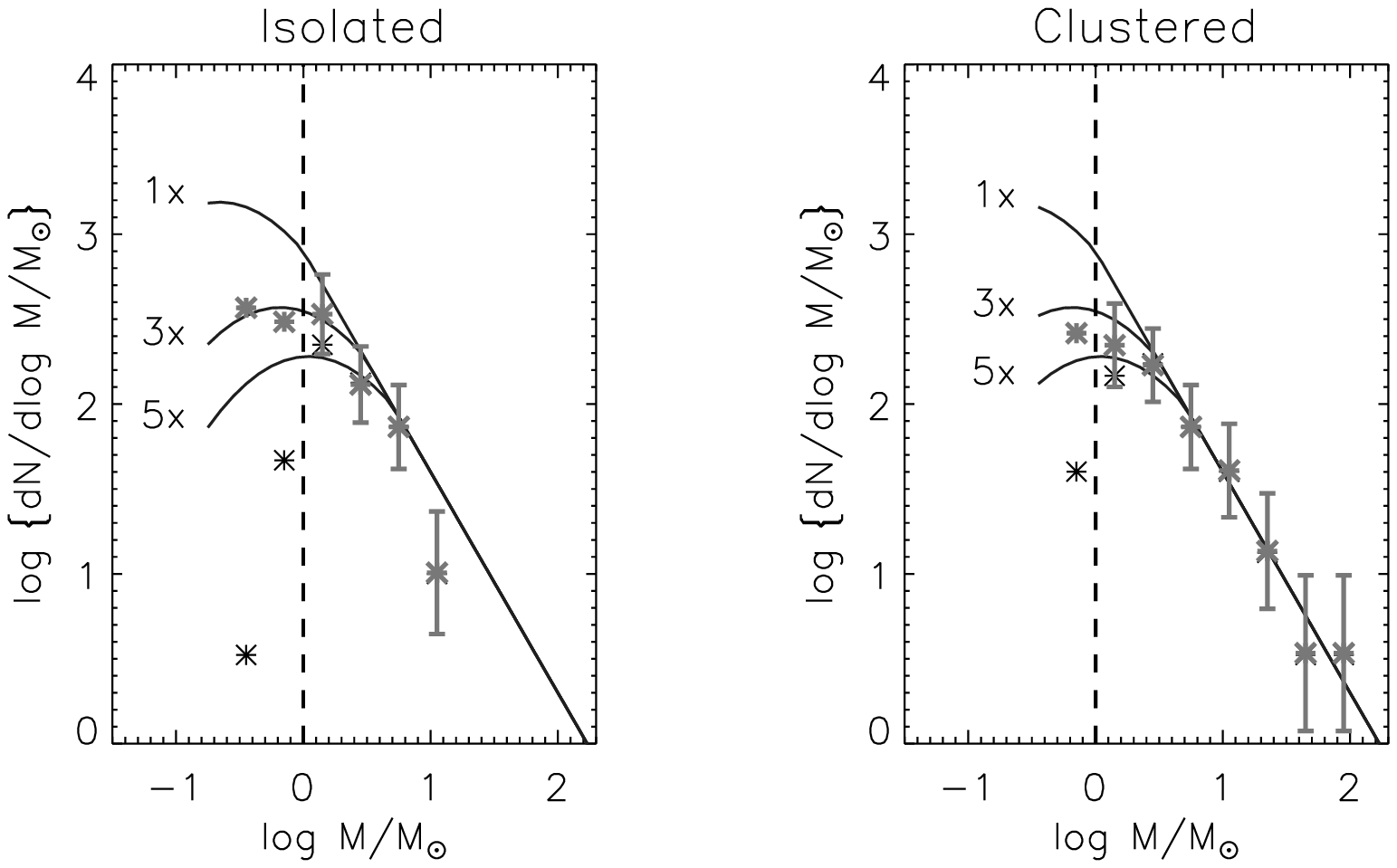}
    \caption{Completeness-corrected CMFs for cores in the shallow survey catalog split at the median N=4 nearest neighbour distance.  Points and overlays follow the convention established in the top row of Fig.~\ref{fig:rawcmfs}.  The isolated core CMF exhibits a clear deficit at  $>$10~$M_{\odot}$ relative to the smooth power law behavior of the clustered core CMF. Otherwise, the two CMFs agree well.}
    \label{fig:cmfclust}
\end{figure}

First, we explore whether core clustering has any impact on the observed CMF.  Since relative spatial completeness is required to probe a range of clustering environments, we restrict this analysis to the shallow survey cores.  We adopt the simple, empirical approach of dividing the shallow core sample at their median $N=4$ nearest neighbour spacing of 166.4\arcsec, or 0.66~pc, with {\it clustered} cores below that value and {\it isolated} cores above it.  The split CMFs are shown in Fig~\ref{fig:cmfclust}, and exhibit strong similarity at low masses.  The high-mass end of the isolated core CMF is deficient relative to the clustered core CMF, however.  High mass cores in our survey are the explicit domain of cluster-forming clumps in the Mon~R2 cloud.  This could be an intrinsic property of clustered cores \citep[e.g.][]{Myers2009}. Of course, one caveat is the risk that cores could be blended together in our data increases as core spacing shrinks.  Higher angular resolution continuum data and line emission mapping to disentangle line of sight blends is necessary to settle this interpretive dichotomy \citep[e.g.][]{Walsh2007}.  

\begin{figure}
	\includegraphics[width=\columnwidth]{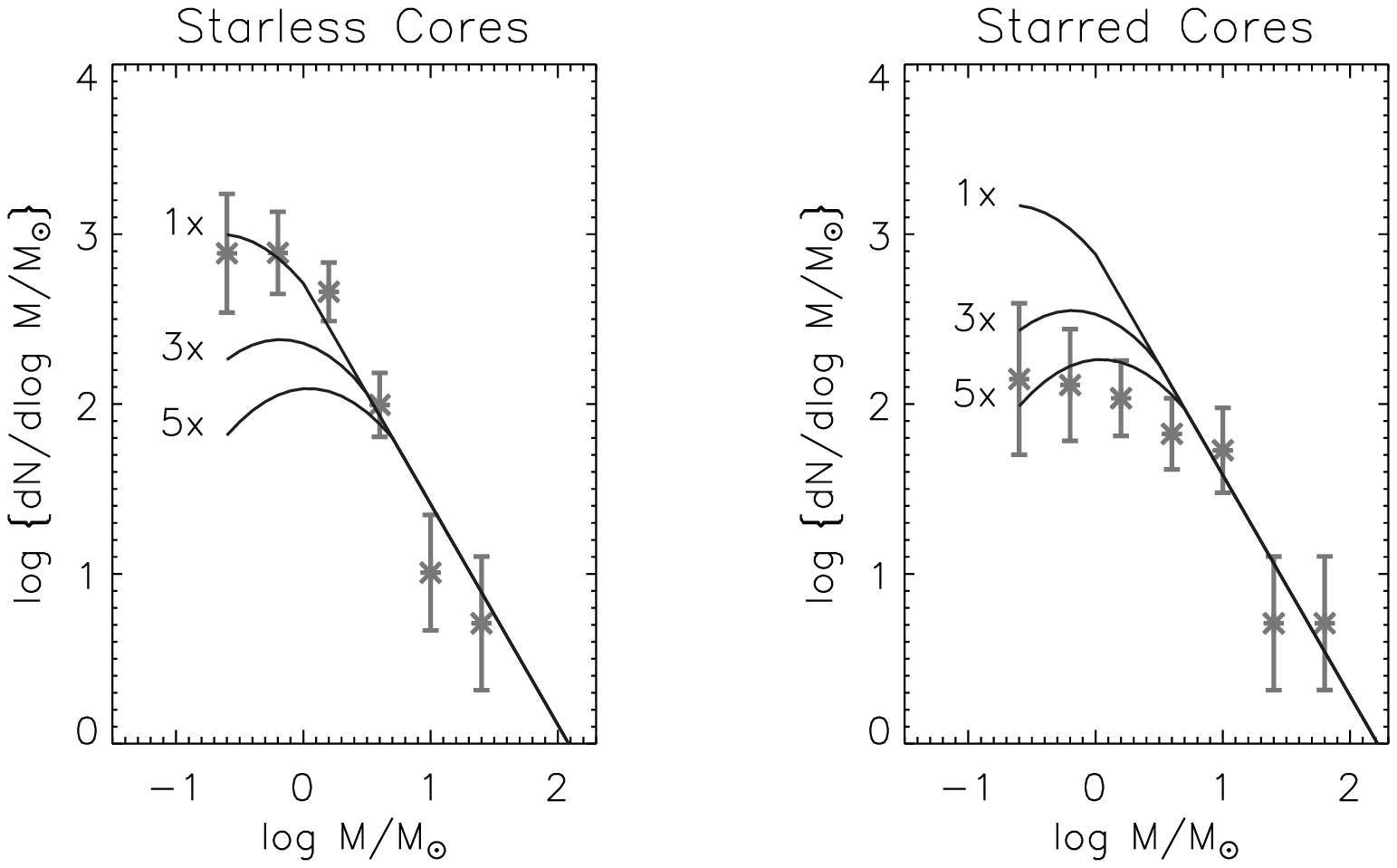}
    \caption{Completeness-corrected CMFs for cores in the shallow and deep surveys, scaled and merged as in Fig.~\ref{fig:rawcmfs}, and split by the presence of a Spitzer-identified YSO within each core's emission footprint.  Plot points and overlays follow the convention established in the combined CMF for all cores in Fig.~\ref{fig:rawcmfs}.  Both CMFs are consistent with a Salpeter power law at masses $>$5~$M_{\odot}$, but the starless core CMF turns over at considerably lower mass (1 vs 5~$M_{\odot}$) relative to the cores with protostars.}
    \label{fig:cmfstarless}
\end{figure}

Next, we examine the CMF of starless cores and cores with protostars.  Despite some differences in YSO sensitivity in more crowded and nebulous environments \citep[e.g.][]{Gutermuth2015}, we expect the typical sensitivity of embedded YSOs among both the deep and shallow surveys to be relatively similar on average.  Thus we adopt the merged CMF for this part of the analysis.  The split CMFs are plotted in Fig.~\ref{fig:cmfstarless}.  Unlike in Fig.~\ref{fig:cmfclust}, where the subsets have equal numbers of members, here the subsets are not equal (209 shallow and 65 deep cores are starless; 86 shallow and 14 deep cores have YSO counterparts), thus the set of shifted and scaled Chabrier IMFs overlaid in Fig.~\ref{fig:cmfstarless} are slightly offset to better match each CMF for $\rm log( M/M_{\odot}) > 0.3$.  As with the clustered analysis, one portion of the pair of CMFs is quite consistent.  However, in this case, it is the power law character of the high-mass end of the CMFs ($>$5~$M_{\odot}$) that agrees.  Below that point, the protostellar core CMF clearly turns over, while the starless core CMF continues to rise until a possible turnover at $\sim$1~$M_{\odot}$, although we cannot rule out an unbroken power law with these data.  

Based on the relatively narrow range of core sizes in Mon~R2 (Fig.~\ref{fig:massvsize}),the presumed reason for the substantial difference in the low mass end of the starless and protostellar CMFs is the growing fraction of cores below the BE stability criterion for masses below 5~$M_{\odot}$.  Meanwhile, the lack of difference in the 1-5~$M_{\odot}$ CMFs for clustered and isolated cores suggests that these seemingly stable cores are formed in both diffuse and dense environments.  As noted by \cite{Myers2009}, the birth site of a core may have a profound effect on its accretion from the ambient diffuse gas environment.  If core clustering is correlated with the density of larger scale cloud structure, then we should expect confinement of high mass cores to denser subregions in the cloud.  We explore this possibility in the next section.


\section{Relationship between cores and natal gas morphology}
\label{sec:cfe}

While it is well established that stars form in dense cores, less is understood about core physical properties and early evolutionary stages. Understanding the physical nature and onset of core formation sheds light on ambiguities such as how many stars each core forms, the initial distribution of core masses, and whether stars are frequently ejected from their birth sites. The details behind these processes set the timescale for star formation and determine how efficiently \textit{diffuse} gas mass is converted into \textit{dense} core mass. 


Studies of Mon~R2 provide an opportunity to understand the role of core formation as a precursor to star formation in a cloud with a wide range of physical conditions. Thus far, many efforts have been made to study cores in nearby molecular clouds \citep{Johnstone2004, Hatchell2005, Johnstone2006a, Johnstone2006b, Young2006, Belloche2011, Kirk2016, Johnstone2017}. For example,  previous surveys have mapped 7.5 deg$^{2}$ (140~pc$^{2}$ at d= 250~pc) in Perseus, 10.8 deg$^{2}$ (50~pc$^{2}$ at d= 125~pc) in Ophiuchus, and 1.5 deg$^{2}$ (30~pc$^{2}$ at d= 260~pc) in Serpens, detecting 122, 44, and 25 cores, respectively \citep{Enoch2006, Enoch2007}. In this survey, we map 2 deg$^{2}$ of Mon~R2 (210~pc$^{2}$ at d=830~pc) and detect 295 cores. Not only does Mon~R2 have an abundance of 1.1~mm cores distributed throughout both high and low column densities, but its active centre harbors intermediate mass protostars that radiate $\sim 10^{4}$~L$_{\odot}$ and drive large outflows \citep{Maaskant}. Recent mid-IR studies of Mon~R2 by \citet{Gutermuth2011} indicate a homogeneous distribution of $\sim$ 1000 Class I \& II young stellar objects (YSOs) with excess infrared emission. They define Class I YSOs as protostars and Class II YSOs as pre-main sequence stars with disks. \citet{Gutermuth2011} also show a clean power-law correlation between YSO surface density and gas column density in Mon~R2, indicating denser concentrations of YSOs coincident with the denser regions of the cloud.  This result suggests that local gas column density dictates stellar spatial distribution, motivating the present study of surface densities of the dense gas cores that represent the preceding stage of star formation.
	
With a statistically significant sample of the core population in Mon~R2, we 
assess the range of core formation environments across the Mon~R2 cloud.  Using the LMT/AzTEC shallow survey cores list and the Herschel-derived gas column densities \citep{Pokhrel2016}, we explore how gas column density sets the core clustering and formation efficiency at both local, parsec scales and integrated over the bulk of the cloud area. 


\subsection{A local correlation between core clustering and gas column density}


To quantify core clustering in Mon~R2, we performed an $n^{th}$ nearest neighbour surface-density analysis, similar to the analysis in \citet{Gutermuth2011}.  Nearest neighbour distances (hereafter, NND) are measured from each core to its $n^{th}$ nearest neighbour. Surface density measurements centred on each core around its NND locality are calculated as $(n-1)/ \pi(NND_{n})^{2}$. Thus, a core in a less crowded region will have its surface density calculated over a much larger area than a core with many close neighbours, though each area will contain the same number of cores. To derive core mass density, we multiply by the mean mass of the core's nearest $n$ neighbours within the enclosed region. We correct the masses as per our correction method in $\S$\ref{sec:observations}. 

To characterize the spatial distribution of gas, we use a Herschel-SPIRE derived gas column density map \citep{Pokhrel2016}. The gas column density around each core is measured over the identical size scale as the local core mass density measurements. We take an average of the column density at every pixel within the region defined by the core's nearest neighbour distance.  To assess clustering at different local size scales, we evaluate core mass surface density, $\Sigma_{core}$, and the surrounding gas column density, $\Sigma_{gas}$, at $n=4, 6, 11,$ and $18$ nearest neighbours. Thus,  $\Sigma_{core}$ represents the surface density of all cores within the NND$_{n}$ area.

 \begin{figure*}
 
 \centering
\minipage{0.32\textwidth}
  \includegraphics[scale=.47]{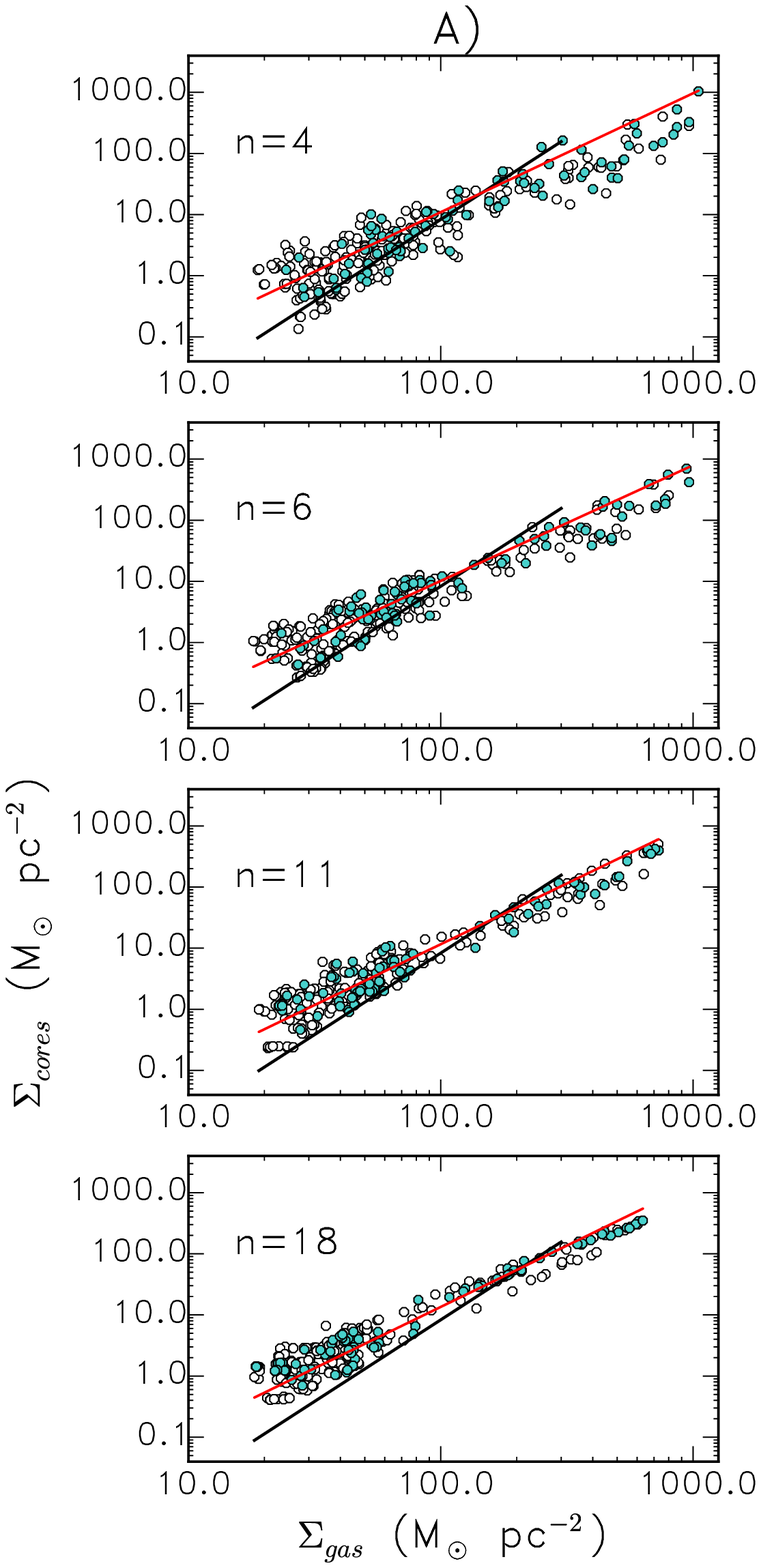}

\endminipage\hfill
\minipage{0.32\textwidth}
  \includegraphics[scale=.47]{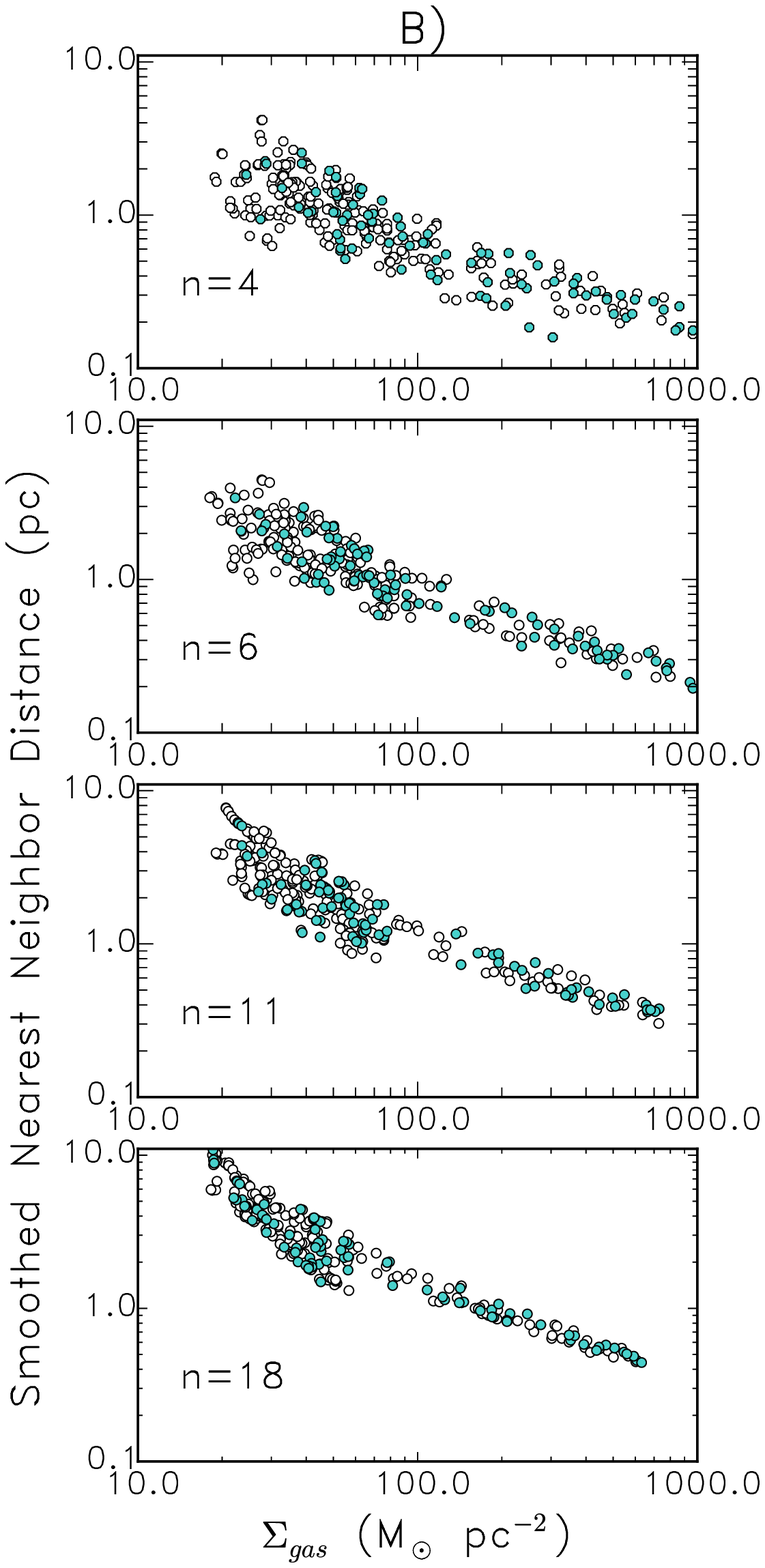}

\endminipage\hfill
\minipage{0.32\textwidth}%
  \includegraphics[scale=.47]{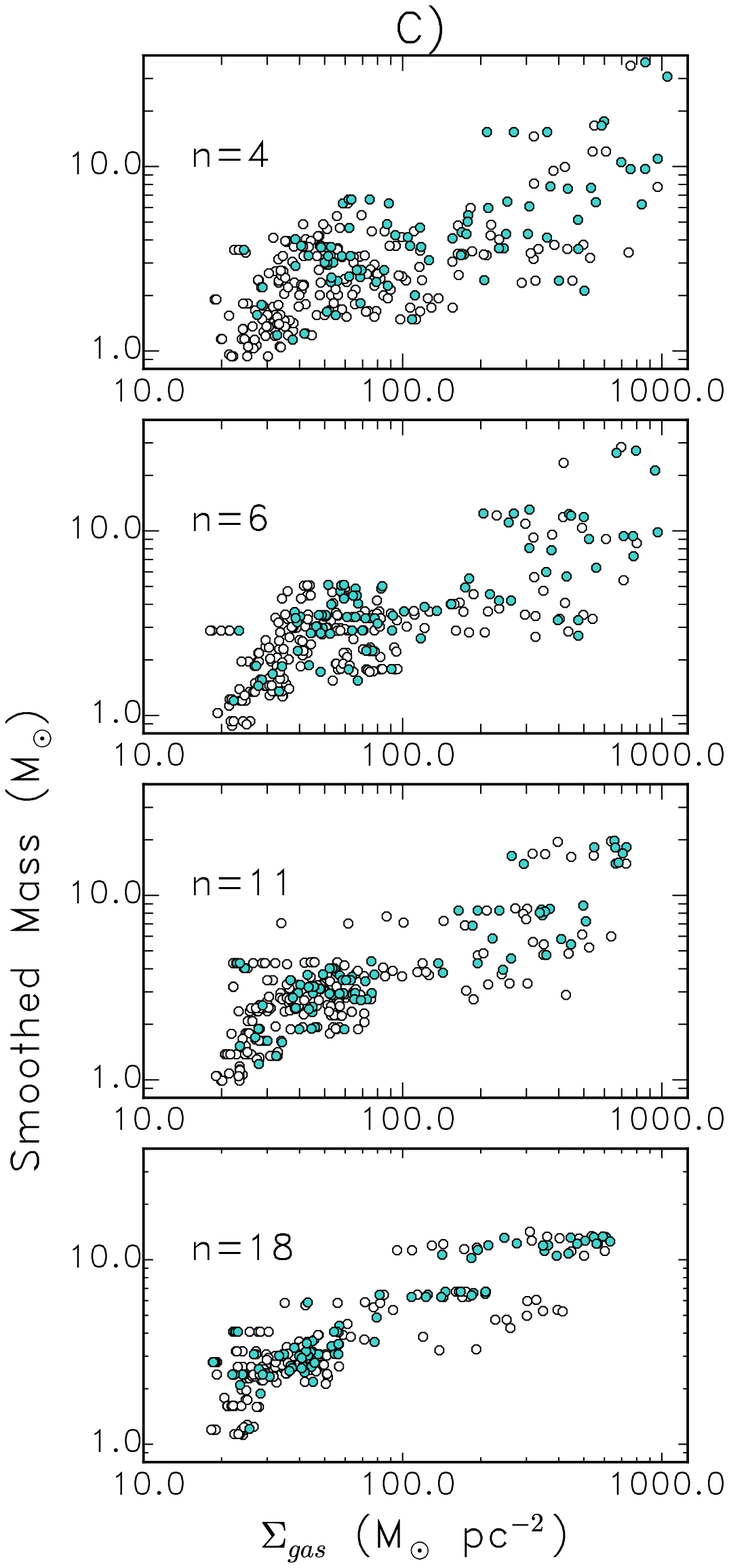}

\endminipage
\caption{Plots demonstrating systematic effects on locally measured core and diffuse gas column densities as a function of $n$, the number of neighbouring cores per measurement.  In each column, measurements centred on starless cores are shown as open circles and those
centred on cores with protostars are filled in cyan circles. 
Column A) the core-gas correlation, shows the smoothed core mass surface density versus smoothed diffuse gas column density at varying $n$ values. The red line shows the power-law fit to the correlation. The black line shows the fit to the star-gas correlation in Mon~R2 \citep{Gutermuth2011} with a power-law index of 2.67. 
Column B) the nearest neighbour distances, thus the smoothing size scales, for the core-gas correlation data in Column A versus smoothed Herschel gas column density. 
Column C) the mean smoothed core mass plotted versus smoothed column density.}

\label{fig:smoothing}
\end{figure*}

In Figure \ref{fig:smoothing}a we plot the core-gas correlation, $\Sigma_{core}$ versus $\Sigma_{gas}$ for $n=4, 6, 11,$ and $18$ nearest neighbour clustering analysis. The best fit power-law index to the core-gas correlation at $n=11$ is $1.99 \pm 0.03$.  The slopes do not change significantly when varying smoothing size scale, where the additional best fit slopes are $1.95 \pm 0.03$ for $n=4$, $1.89 \pm 0.03$ for $n=6$ and $2.00 \pm 0.03$ for $n=18$.  The black line indicates the power-law fit to the star-gas correlation in \citet{Gutermuth2011}. The star-gas correlation describes YSO mass density as a function of local gas column density as traced by near-IR extinction. A constant YSO mass of $M_{\text{YSO}}= 0.5  M_{\odot}$ was adopted \citep[mean mass for a fully sampled IMF;][]{chabrier2005}. The best-fit power-law index to the star-gas correlation is $2.67 \pm 0.02$, with a smoothing size scale of $n=11$ and an identical $n^{th}$ nearest neighbour surface density analysis. 

In Figure \ref{fig:smoothing}b we plot the nearest neighbour distances versus the corresponding smoothed column density for each $n$ value. Nearest neighbour distances used as a smoothing size scale range from 0.2~pc (n=4) to 10.0~pc (n=18). We find that regions of the highest column density are contained within the smallest nearest neighbour distances. Thus, higher column density regions are observed to be more densely populated with cores. The correlation becomes tighter and the range in $NND_{n}$ increases as we increase the measurement area to include more neighbours. 

Figure \ref{fig:smoothing}c shows the mean core mass (as opposed to the mean core mass per area in Figure~\ref{fig:smoothing}a) enclosed at varying $n$. We find that cores at higher column densities have a higher mean mass than cores at lower column densities.  However, this mass difference could be the result of reaching the spatial resolution limits of the survey.  Since cores appear more densely clustered at higher column density, core blending may become more common at higher column densities, resulting in a higher apparent mean mass. Accounting for local mass variation ensures that regardless of the reason for the apparent mass bias with gas column density, our core-gas correlation measurements should be unbiased by this effect.

\subsection{Thermal fragmentation scenario of star formation}

A similarity between YSO positions and natal molecular gas structure is found in many star forming regions \citep[e.g.][]{Allen2007,Gutermuth2009}.  \citet{Gutermuth2011} explored this relation further, showing that Spitzer-identified YSOs in eight nearby molecular clouds, including Mon~R2, appear to cluster in structures similar to the structure of the natal molecular gas. Each cloud exhibits a positive power law trend between stellar surface density and local gas column density at parsec scales. The correlations observed are all super-linear power laws with indexes ranging from 1.37-3.8.  Spreads in each cloud's star-gas locus are correlated with relative evolutionary differences among the YSOs.  



\citet{Gutermuth2011} presented a simple model to explain the observed star-gas correlation and its evolution in a cloud.  Under the assumption of little motion between YSOs and gas at parsec scales, thermal fragmentation of an isothermal, self-gravitating, modulated layer of gas can produce a star formation rate (SFR) gas column density power law of index 2 within a cloud \citep{Myers2009}.  As stars are forming according to this law, the stellar mass surface density rises in all locations.  While the underlying gas column density is not significantly depleted, the instantaneous star-gas correlation preserves the power law index of the SFR-gas density law ($\sim$ 2).  For any power law of super-linear index, stars are formed more efficiently at higher column densities.  Thus regions of relatively high gas column density experience significant gas mass depletion before those at lower gas column densities, resulting in a net steepening of the observed star-gas power law with time.  

\begin{figure}
\centering
	\includegraphics[width=\columnwidth]{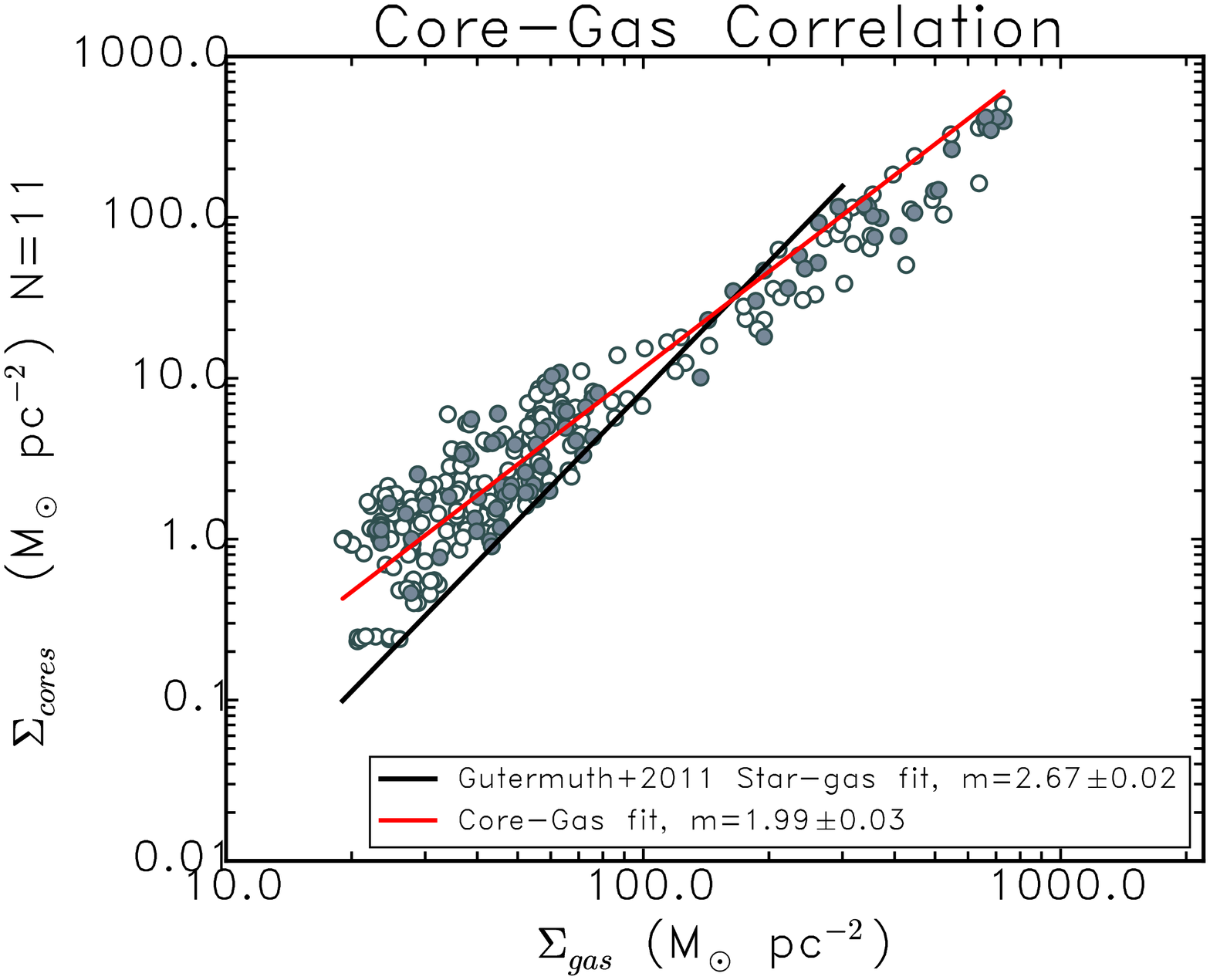}
    \caption{The core-gas correlation plotted smoothed over a size scale corresponding to $n=11$ nearest neighbours. The filled grey circles represent cores with protostars (overlapping Spitzer-identified YSOs) and open circles represent starless cores without IR excess.  The red line shows the power-law fit to the core-gas correlation with a best-fit power-law index of $1.99 \pm 0.03$. The black line indicates the star-gas correlation in \citet{Gutermuth2011} with a power-law index of $ 2.67 \pm 0.02$.}
    \label{fig:sgcg}
\end{figure}

In Figure~\ref{fig:sgcg}, 
we show the core-gas correlation measurements for $n=11$ nearest neighbour smoothing with the best-fit power-law of index of 1.99 (red), in contrast to the star-gas power law index of 2.67 (black). Our observed core-gas correlation slope is consistent with the thermal fragmentation model prediction, effectively tracing the SFR-gas law itself.  Meanwhile, the steeper star-gas power law is consistent with the stellar buildup and local gas mass depletion at higher column density locations, as described in \citet{Gutermuth2011}.  If the primordial gas distribution has been somewhat depleted at high column densities, then the agreement between the core-gas correlation and the thermal fragmentation model suggests that the core formation and evolution timescale is relatively rapid in these regions. Additionally,  the current core properties seem to be dictated by recent changes in the gas column density environment of the denser regions.


\subsection{Core formation efficiency at global scales}

We report core formation efficiencies (CFEs) sampled from the entire cloud and categorized by $A_{V}$ bin. CFE is a measure of the fraction of the mass that resides within cores relative to the total mass over a given $A_{V}$ interval. By integrating our measurements over the entire cloud, we obtain a global quantification of core formation efficiency to assess whether core formation is directly influenced by natal gas morphology or behaves uniformly throughout the cloud at similar $A_{V}$. 

\begin{figure}
\centering
	\includegraphics[width=\columnwidth]{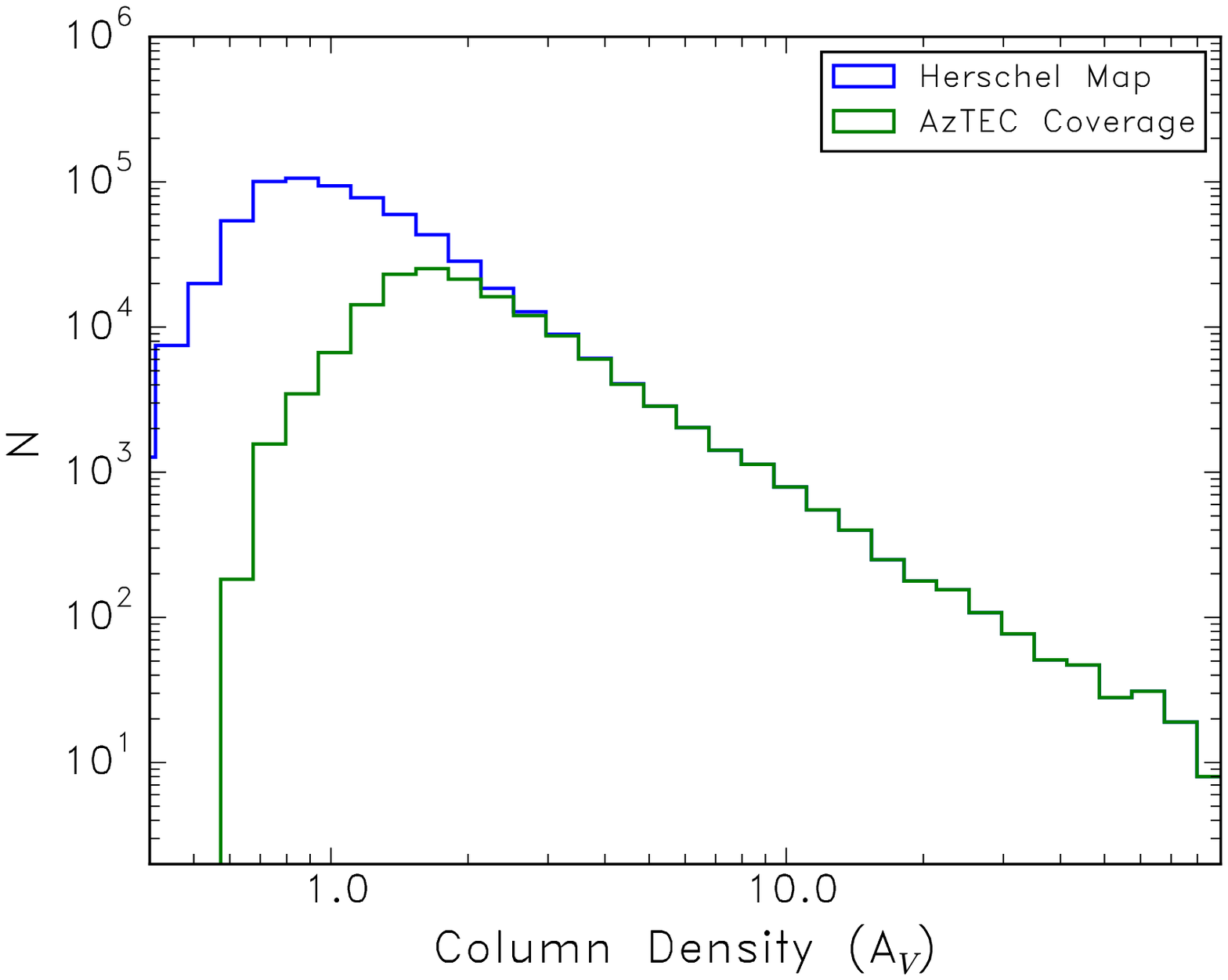}
    \caption{Column density histograms of Mon~R2 cloud coverage in the entire Herschel map (blue) and the portion covered by the LMT/AzTEC survey (green). The AzTEC survey, guided by the Herschel map, covers high density regions rich in substructure. The green AzTEC histogram deviates from the Herschel histogram near $A_{V} \sim 3$, where there is less contribution from low-column density regions. The lowest column density probed with AzTEC (regions mapped in Figure~\ref{fig:coverage}) is $\sim$0.6 $A_{V}$.  }
    \label{fig:histo_cov}
\end{figure}

The LMT/AzTEC shallow survey (coverage contours plotted in Figure \ref{fig:coverage}) covers the regions of Mon~R2 that are rich in substructure but it is blind to emptier low column density regions mapped by Herschel \citep{Pokhrel2016}. Figure \ref{fig:histo_cov} shows the column density histograms for the entire Herschel map (blue) and the AzTEC survey coverage within (green).  In the more diffuse regions of the cloud, below $A_{V} \sim 2$, the spatial limits of the AzTEC survey result in a considerably lower fraction the cloud being surveyed. Since our cores were identified using AzTEC, we only sample from gas column densities in the portions of the Herschel map covered by AzTEC. 

 \begin{figure}
\centering
	\includegraphics[width=\columnwidth]{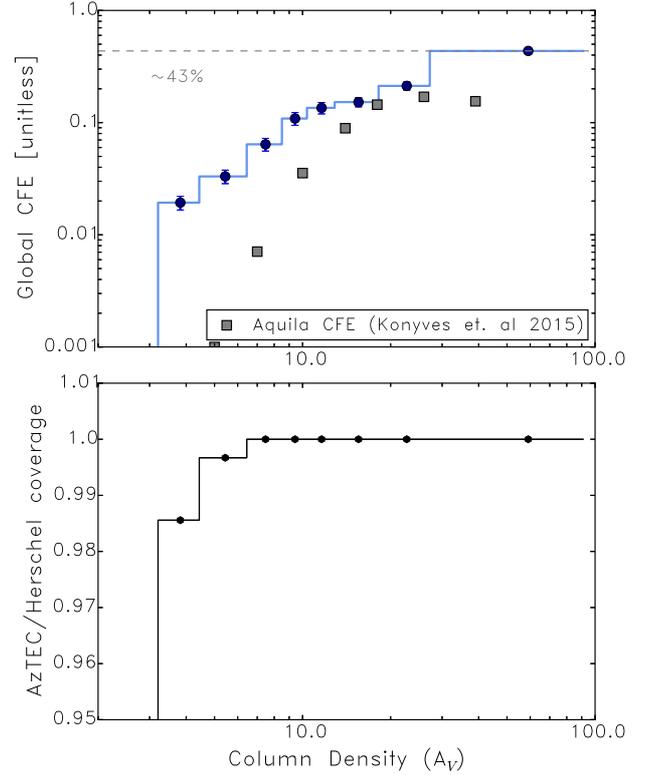}
    \caption{Observed global core formation efficiency (CFE) as a function of background column density (top panel) with the AzTEC to Herschel coverage ratio in the bottom panel at corresponding $A_{V}$ bins. We follow \citet{Konyves15}  and define $\text{CFE}_{\text{global}}= \Delta M_{\text{cores}} (A_{\text{v}}) / \Delta M_{\text{cloud}} (A_{\text{v}}) $, where the CFE error is propagated summation of the uncertainties associated with core mass corrections (numerator) and uncertainties in the Herschel column density map (denominator).  The bottom panel indicates that, within the $A_{V}$ bins we estimate core formation efficiencies, we still achieve a global column density measurement ($>$98\%) despite the AzTEC preferential coverage of higher column densities. In the CFE plot, the horizontal dashed line indicates the maximum value the efficiency approaches, CFE $\sim 43\%$ at $A_{V} > 15$. We overplot the CFE results in \citet{Konyves15} for cores in Aquila (grey squares), with a trend that asymptotically approaches CFE$\sim 15\%$  at $A_{V} > 15$. }
    \label{fig:cfe}
\end{figure}


In Figure \ref{fig:cfe} we follow the analysis of \citet{Konyves15} and plot the core formation efficiency versus column density. As in \citet{Konyves15} we define the observed core formation efficiency as $\text{CFE}_{\text{global}}= \Delta M_{\text{cores}} (A_{\text{v}}) / \Delta M_{\text{cloud}} (A_{\text{v}}) $ where $\Delta M_{\text{cores}} (A_{\text{v}})$ is the mass of cores  that lie within a given bin of background $A_{\text{v}}$ values and $\Delta M_{\text{cloud}} (A_{\text{v}})$ is the cloud mass estimated from the Herschel column density map in the same $A_{\text{v}}$ bin. The column density bin ranges are generated such that each bin contains 30 cores. In the bottom panel of Figure~\ref{fig:cfe} we plot the ratio of mass in AzTEC coverage to total mass in the Herschel map within the identical $A_{\text{v}}$ bins. Despite the lack of coverage in very diffuse low column density areas, even at the smallest $A_{V}$ bin shown (  3-5 $A_{V}$) the maps sample $98.5\%$ of the cloud mass.

\begin{figure}
\centering
	\includegraphics[width=\columnwidth]{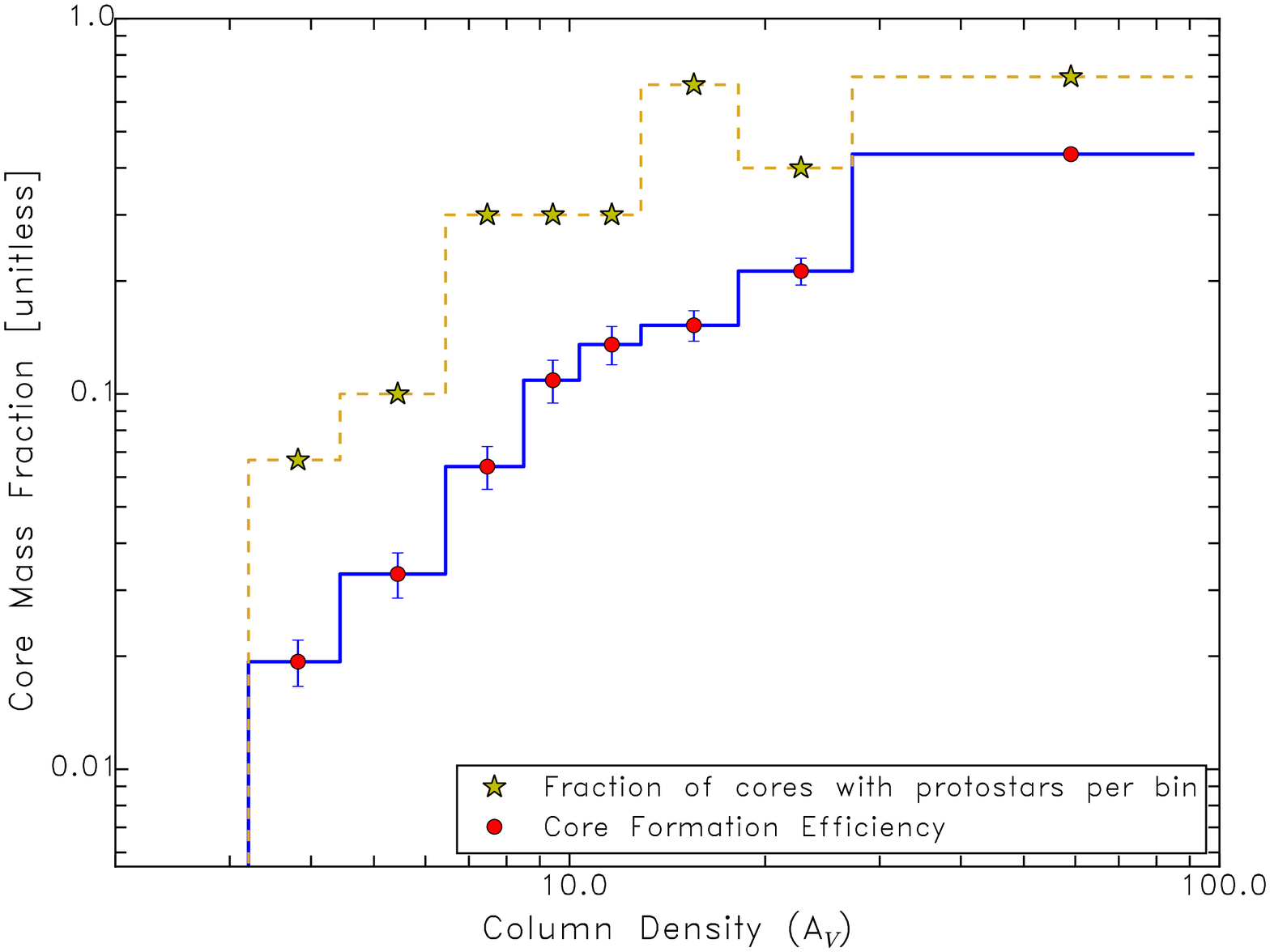}
    \caption{The global core formation efficiency as a function of column density $A_{V}$ bin (as shown in Figure~\ref{fig:cfe}) plotted along with the fraction of cores with protostars per bin as yellow stars. The yellow dotted line represents the corresponding bin width, where each bin contains 30 cores. }
    \label{fig:starredfrac}
\end{figure}

We find that the observed global CFE rises with higher column density and approaches $\sim$ 43\% efficiency for the highest column density bin above  $A_{\text{v}} \sim 30$. \citet{Konyves15} found that the core formation efficiency for cores in Aquila asymptotically approaches $\sim15\%$ when $A_{\text{v}} > 15$. We plot their CFE values in Figure \ref{fig:cfe} (grey squares). The two trends approximately meet at CFE=15$\%$ where  $A_{\text{v}} \sim 15$; however, Mon~R2's efficiency rises and does not level off like Aquila's. Cores in \citet{Konyves15} are identified using Herschel-SPIRE 500$\mu$m data at 36\arcsec resolution. Though AzTEC has better angular resolution, Aquila is significantly closer. Therefore, the corresponding physical beam sizes in the two studies are comparable.  However, the uncertainty in Aquila's adopted distance could give rise to such beam size effects making it difficult to separate cores from larger cloud structure. In this case the varying CFEs might be a result of more detectable cores in MonR2. Additionally, the differences between the observed CFE trends in these two clouds may be attributed to evolutionary differences that modify the cloud's gas reservoir abundance and subsequent accretion onto its star-forming cores.


As a follow-up to the core formation efficiency (CFE) analysis, in Figure~\ref{fig:starredfrac} we show the global CFE along with the fraction of cores per bin that have protostars (yellow star markers with dashed line), where each bin contains 30 cores.  Doing so categorizes the percentage of cores that have formed protostars by column density and relates the column density dependence to the formation efficiency.  This fraction increases with higher column density similarly to the CFE. To explain this, we consider a scenario in which cores are prohibited from forming until their filaments become super-critical. This process happens more rapidly in regions of higher mean column density due to the higher abundance of diffuse gas that is readily available for mass accretion compared to low density regions of the cloud. The Bonnor-Ebert condition maintains that cores require a certain density to become unstable to further collapse. This happens more effectively and with a higher rate in high column density regions, thus accounting for a higher fraction of cores with protostars in high column density regions of the cloud, as observed.

\subsection{Local vs. Global Core Formation Efficiencies}
  \begin{figure}
\centering
	\includegraphics[width=\columnwidth]{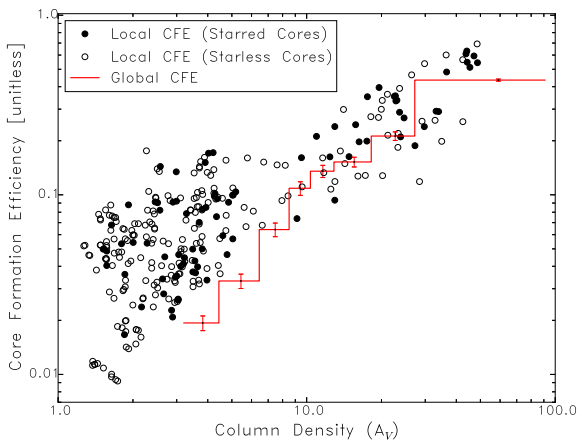}
    \caption{The global CFE (shown in Figure~\ref{fig:cfe}) plotted again (in red) with local CFE measurements for each core shown as black circles, where filled circles represent cores with protostars and open circles represent starless cores. We define local CFE=$\Sigma_{core}/\Sigma_{gas}$ around $n=11$ nearest neighbours, where $\Sigma_{core}$ is the mean core mass density and $\Sigma_{gas}$ is the mean Herschel column density smoothed over each core's 11th nearest neighbour distance. Below $A_{V} \sim 10$, local CFE is higher than global CFE. }
    \label{fig:localglobal}
\end{figure}

As a function of $A_{\text{v}}$ bin, our CFE indicates a lower efficiency at low column density when the core mass and cloud mass are sampled uniformly throughout the entire cloud. How would this efficiency change if it were measured on a local scale? In Figure \ref{fig:localglobal} we plot the global CFE again with a local measurement of core formation efficiency. We define $CFE_{\text{local}}= \Sigma_{\text{cores}}/ \Sigma_{\text{gas}} $ at $n=11$ and plot it versus local n=11 $\Sigma_{\text{gas}}$. We follow the convention that filled circles represent density measurements around cores with protostars, and circles represent density measured around starless cores. For $A_{\text{v}} < 7$, the locally measured CFE is higher than the global CFE by about a factor of 4, while they agree nicely for $A_{\text{v}} > 7$.

While both local and global trends, as well as \citet{Konyves15}, indicate lower core formation efficiency at lower column densities, it is not clear whether this is because cores form less efficiently when surrounded by low column density gas or that lower column density regions contain lower mass cores that are below our detection limits. One possibility is that the observed difference between locally and globally measured CFEs is due to the fact that lower column density bins contain fewer cores, contributing more gas mass to the denominator of the global CFE calculation and thus reducing it. We also speculate that projection effects play a role, such that higher column density regions trace high volume density regions better than low column density regions do \citep{Krumholz2012}. For example, a given column density can be achieved through a combination of elevated volume densities and longer lines of sight, resulting in perceived lower global CFE for cores lie in low column density regions. Therefore, locally sampled points will be biased toward places with truly elevated column density rather than simply long lines of sight.

\section{Summary}
\label{sec:summary}

We use the AzTEC instrument on the Large Millimeter Telescope (LMT) to obtain an extensive census of 1.1~mm dense gas cores in the Monoceros R2 molecular cloud complex. We also use Herschel SPIRE 500$\mu$m data to probe large scale gas morphology and determine the relationship between dense cores and their gas environment. The Mon R2 cloud is 830~pc away and spans a size of 40~pc $ \times$ 40~pc. We summarize the main results below.

\begin{enumerate}

\item In the $\sim$ 2 deg$^{2}$ shallow LMT/AzTEC survey of Mon~R2 we identify a total of 295 1.1~mm cores, 86 with protostars and 209 starless.  We detect 79 cores in the additional deep survey, of which 65 are starless and 14 have protostars. Cores with protostars are 1.1~mm sources with overlapping Spitzer-identified YSOs, while starless cores may or may not eventually form stars. With its $8\arcsec$ resolution, LMT/AzTEC (at 32m) works well as a core-finder in the relatively distant Mon~R2 cloud (d=830~pc), as  the physical beam size of $\sim$0.03~pc matches the typical dense core size of $\sim 0.05 \times 0.05$~pc. 

\item We correct for the observed systematic effects of a shallow (7~mJy/beam RMS) survey by scaling up under-detected fluxes and sizes for cores based on their observed peak-to-total flux ratio as a function of their S/N.  Adopting the corrected values, the cores have a median mass $\sim 2.1 M_{\odot}$ and a median deconvolved FWHM size $\sim 0.08$~pc. We plot the mass versus size for all cores and find that $\sim 44\%$ of cores lie above the Bonnor-Ebert stability line for cores with $T=12K$, indicating that their density structure makes them unstable to gravitational collapse in the absence of any non-thermal support. $\sim 90\%$ of the cores below the line are starless.

\item We construct the core mass function of both the shallow and deep survey cores, correcting for core incompleteness and residual false core contamination, and split based on core clustering and the presence or lack of one or more embedded young stars.  We demonstrate that the overall CMF on Mon~R2 was broadly consistent in shape with those reported for other clouds.  We found that high mass cores ($>$10~$M_{\odot}$) are exclusively found among denser clusters of cores, although core blending could play a role.  In addition, we found that the CMF for cores with protostars turns over at much higher masses ($\sim$5~$M_{\odot}$) than that of starless cores ($\sim$1~$M_{\odot}$).  Starless cores vastly outnumber cores with protostars at lower masses.

\item We present a core-gas correlation to locally estimate the relationship between core clustering and surrounding gas column density at $\sim$pc scales. We find a power law relationship with an index of $1.99 \pm 0.03$ using an $n=11$ nearest neighbour distance smoothing size scale. The slope is consistent with the star-forming model of thermal Jeans fragmentation in an isothermal self-gravitating layer predicted in \citet{Gutermuth2011}. Given the star-gas correlation slope of $2.67 \pm 0.02$  \citep{Gutermuth2011} in Mon~R2, we postulate that gas depletion upon YSO formation steepens the primordial (core-gas) power-law. Thus, gas dispersal in high column density regions may occur earlier than low column density gas depletion and produce an over-density of stars relative to nearby gas, resulting in a steeper star-gas power-law. 

\item We compute a global core formation efficiency sampled from different $A_{\text{v}}$ bins in the cloud and find lower core formation efficiencies at lower column densities, with a maximum CFE of $\sim$ 43$\%$ above $A_{\text{v}} \sim 30$. We compare this global CFE with a local CFE defined as the ratio between $\Sigma_{core}$ and $\Sigma_{gas}$ at $n=11$.  Below  $A_{\text{v}} \sim 7$, the local CFE is greater than the global CFE by a factor of $\sim$ 4, which we propose could be an effect of varying lines of sight and elevated volume densities to produce low column densities, whereas high column densities trace high volume densities more efficiently.  We also show that the fraction of cores with protostars in each bin increases with higher column density.  We suggest that in these regions there is more gas available for accretion onto cores and filaments.  This may expedite a core's advancement to the critical density required for gravitational collapse, causing a higher fraction of cores with protostars in high column density portions of the cloud. 
\end{enumerate}

\section*{Acknowledgements}

We thank the referee for their helpful suggestions that improved this paper. ADS was supported by the NEAGAP/IMSD Fellowship at the University of Massachusetts Amherst during this project.  RAG's participation in this project was supported by NASA ADAP grants NNX11AD14G, NNX15AF05G, and NNX17AF24G, NASA JPL/Caltech contract 1489384, and NSF grant AST 1636621 in support of TolTEC, the next generation mm-wave camera for LMT.  The AzTEC instrument was built and operated through support from NSF grant 0504852 to the Five College Radio Astronomy Observatory.  The authors gratefully acknowledge the many contributions of David Hughes in leading the LMT to its successful operational state.  The authors also thank the personnel and observers at the LMT that helped obtain the observations, including Victor Gomez, Ricardo Chavez, Alfredo Monta$\tilde{\rm n}$a, Erica Keller, James Lowenthal, Jes\'us Rivera, Mihwa Han, Miguel Ch\'avez, Fabian Rosales, Jorge Zavala, Salvador Ventura, Luis González, Miguel Velazquez, Edgar Castillo, Emmaly Aguilar, Daniel Rosa, Ra\'ul Maldonado, Milagros Zeballos, and Yuping Tang.  This work is based on observations made with Herschel, a European Space Agency cornerstone mission with science instruments provided by European-led Principal Investigator consortia and with significant participation by NASA. This work is based in part on observations made with the Spitzer Space Telescope, which is operated by the Jet Propulsion Laboratory, California Institute of Technology, under a contract with NASA. 





\begin{table*}

\caption{LMT/AzTEC Observations Table}
\label{table:obs}
\centering
\begin{tabular}{l c c c c c c} 
\hline
Field Name & Centre RA & Centre Dec & Cov. Area & Median Noise & Obs. Dates & Tau (225 GHz) \\
 & (J2000.0) & (J2000.0) & (Sq. Arcmin.) & (mJy/bm) & (YMD) & \\
\hline
\hline 
MonR2-SH-Field01 & 06 08 56.31 & -07 29 13.8 & 798.7 & 4.14 & 2014-12-21 & 0.07 \\ 
  &   &   &   &   & 2015-01-11 & 0.04 \\ 
MonR2-SH-Field02 & 06 05 40.65 & -07 14 43.7 & 346.2 & 5.52 & 2015-01-15 & 0.07 \\ 
MonR2-SH-Field03 & 06 04 33.05 & -07 00 42.0 & 161.2 & 5.98 & 2014-12-20 & 0.09 \\ 
MonR2-SH-Field04 & 06 07 59.89 & -07 00 04.1 & 1319. & 7.30 & 2014-12-05 & 0.07 \\ 
  &   &   &   &   & 2014-12-06 & 0.11 \\ 
MonR2-SH-Field05 & 06 09 19.80 & -06 44 54.2 & 161.2 & 6.82 & 2014-12-20 & 0.09 \\ 
MonR2-SH-Field06 & 06 04 05.05 & -06 38 31.4 & 133.6 & 3.55 & 2014-12-21 & 0.07 \\ 
  &   &   &   &   & 2015-01-11 & 0.05 \\ 
MonR2-SH-Field07 & 06 05 15.56 & -06 29 41.6 & 342.0 & 4.95 & 2014-12-20 & 0.08 \\ 
  &   &   &   &   & 2014-12-21 & 0.07 \\ 
MonR2-SH-Field08 & 06 15 02.48 & -06 23 45.8 & 280.5 & 3.68 & 2014-12-20 & 0.09 \\ 
  &   &   &   &   & 2015-01-16 & 0.06 \\ 
MonR2-SH-Field09 & 06 07 15.81 & -06 21 33.8 & 2142. & 8.78 & 2014-11-27 & 0.10 \\ 
  &   &   &   &   & 2014-12-02 & 0.03 \\ 
  &   &   &   &   & 2014-12-06 & 0.11 \\ 
  &   &   &   &   & 2015-01-31 & 0.07 \\ 
MonR2-SH-Field10 & 06 12 36.89 & -06 12 11.2 & 484.2 & 8.20 & 2014-12-20 & 0.08 \\ 
MonR2-SH-Field11 & 06 10 58.32 & -06 09 53.3 & 692.8 & 7.04 & 2014-12-10 & 0.07 \\ 
  &   &   &   &   & 2014-12-19 & 0.08 \\ 
MonR2-SH-Field12 & 06 06 01.72 & -06 03 48.8 & 311.8 & 7.48 & 2015-01-31 & 0.03 \\ 
MonR2-SH-Field13 & 06 08 04.69 & -05 47 44.3 & 1240. & 5.07 & 2014-12-11 & 0.06 \\ 
  &   &   &   &   & 2014-12-19 & 0.07 \\ 
  &   &   &   &   & 2015-01-16 & 0.06 \\ 
MonR2-SH-Field14 & 06 07 41.30 & -05 18 34.6 & 735.5 & 7.97 & 2014-12-07 & 0.07 \\ 
  &   &   &   &   & 2014-12-09 & 0.07 \\
MonR2-DP-Map01 & 06 07 40.32 & -06 50 06.0 & 51.59 & 2.24 & 2016-01-25 & 0.03 \\ 
MonR2-DP-Map02 & 06 04 33.12 & -07 01 12.0 & 50.45 & 4.44 & 2016-01-22 & 0.07 \\ 
MonR2-DP-Map03 & 06 07 57.60 & -07 04 12.0 & 49.36 & 2.11 & 2016-02-05 & 0.03 \\ 
  &   &   &   &   & 2016-02-06 & 0.04 \\ 
  &   &   &   &   & 2016-02-18 & 0.04 \\ 
MonR2-DP-Map04 & 06 08 16.80 & -06 58 12.0 & 48.10 & 2.51 & 2016-01-25 & 0.03 \\ 
  &   &   &   &   & 2016-02-05 & 0.03 \\ 
MonR2-DP-Map05 & 06 05 28.80 & -06 29 24.0 & 48.79 & 4.04 & 2016-01-23 & 0.03 \\ 
  &   &   &   &   & 2016-01-24 & 0.02 \\ 
MonR2-DP-Map06 & 06 07 31.20 & -05 09 36.0 & 47.30 & 3.50 & 2016-02-21 & 0.07 \\ 
  &   &   &   &   & 2016-02-25 & 0.09 \\ 
MonR2-DP-Map07 & 06 07 21.60 & -06 45 00.0 & 47.56 & 2.58 & 2016-02-18 & 0.04 \\ 
MonR2-DP-Map08 & 06 04 14.40 & -06 40 48.0 & 51.16 & 6.24 & 2016-01-23 & 0.03 \\ 
MonR2-DP-Map09 & 06 04 52.80 & -06 31 12.0 & 49.51 & 3.27 & 2016-01-22 & 0.04 \\ 
  &   &   &   &   & 2016-01-23 & 0.03 \\ 
  &   &   &   &   & 2016-01-24 & 0.02 \\
\end{tabular}
\end{table*}

\begin{landscape}
\begin{table}

\caption{All core properties. The full version of this table is available online. }
\label{table:allcores}
\centering
\begin{tabular}{l c c c c c c c c c c c c c c c} 
\hline
Core ID & RA & Dec & Area & Tot. Flux & Corr. Flux & $\frac{Corr. Flux}{Tot. Flux}$ &Tot. S/N & Protostar? & HPA & Corr. HPA & Peak Flux & Corr. Mass & Corr. FWHM & PSN & Prob.\\
 & (J2000.0)& (J2000.0) & (Sq. Arcs) & (mJy) & (mJy) & & & (Y/N) & (Sq. Arcs) & (Sq. Arcs) & (Jy Bm$^{-1}$) & ($M_{\odot}$)& (pc) & &  \\
\hline
\hline
1 & 06 08 53.0 & -07 37 29.8 & 259.0 & 22.0 & 46.8 & 2.1 & 4.124 & N & 259.0 & 338.7 & 17.0 & 0.8 & 0.07 & 3.981 & 0.947 \\
2 & 06 08 14.5 & -07 29 26.2 & 344.0 & 26.0 & 55.3 & 2.1 & 4.526 & N & 344.0 & 395.0 & 17.0 & 1.0 & 0.08 & 4.186 & 1.0 \\
3 & 06 09 20.0 & -07 29 1.3 & 462.0 & 36.0 & 75.1 & 2.1 & 5.305 & N & 462.0 & 496.9 & 18.0 & 1.4 & 0.09 & 4.576 & 0.995 \\
4 & 06 09 9.1 & -07 35 15.3 & 741.0 & 55.0 & 146.0 & 2.7 & 6.487 & N & 741.0 & 1022.7 & 16.0 & 2.6 & 0.14 & 4.115 & 1.0 \\
5 & 06 09 29.1 & -07 35 6.7 & 286.0 & 21.0 & 62.4 & 3.0 & 3.952 & N & 286.0 & 528.3 & 14.0 & 1.1 & 0.09 & 3.419 & 0.785 \\
6 & 06 09 11.3 & -07 35 4.8 & 284.0 & 20.0 & 59.3 & 3.0 & 3.85 & N & 284.0 & 503.4 & 14.0 & 1.1 & 0.09 & 3.571 & 0.845 \\
7 & 06 09 31.3 & -07 30 58.2 & 203.0 & 15.0 & 38.2 & 2.5 & 3.445 & N & 203.0 & 335.4 & 14.0 & 0.7 & 0.07 & 3.672 & 0.78 \\
8 & 06 08 50.2 & -07 26 52.8 & 387.0 & 28.0 & 81.4 & 2.9 & 4.491 & N & 387.0 & 633.3 & 15.0 & 1.5 & 0.1 & 3.903 & 0.823 \\
9 & 06 09 9.2 & -07 26 53.8 & 566.0 & 40.0 & 113.8 & 2.8 & 5.511 & N & 566.0 & 862.4 & 15.0 & 2.0 & 0.12 & 4.004 & 0.987 \\
10 & 06 08 47.5 & -07 26 13.8 & 213.0 & 15.0 & 54.8 & 3.7 & 3.249 & N & 213.0 & 501.6 & 13.0 & 1.0 & 0.09 & 3.338 & 0.799 \\

\end{tabular}

\end{table}
\end{landscape}

\bibliographystyle{mnras}
\bibliography{monr2pub}{} 



\appendix

\section{Tests with deep AzTEC data}
\label{sec:appendixtests}

\begin{figure*}
\includegraphics[width=.5\textwidth]{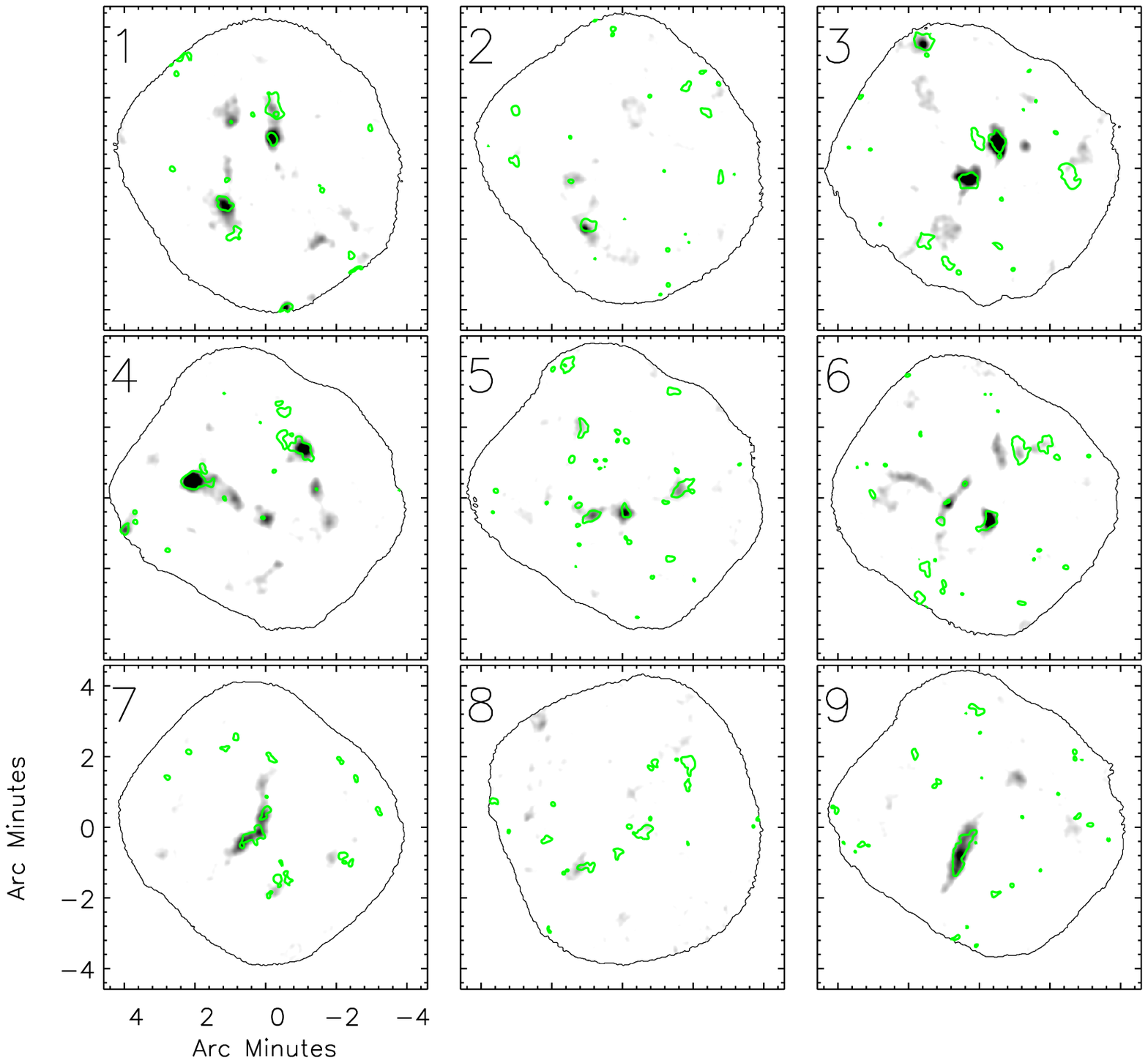}
\caption{Inverted greyscale images of the deep field observations, as shown in the inset of Fig~\ref{fig:aztecdata}, overlaid with S/N=2.5 contours from the shallow survey observations of the same locations.}
\label{fig:deepmapscontours}
\end{figure*}

The deep LMT/AzTEC survey identified a total of 79 cores in 9 fields with masses ranging from 0.07~$M_{\odot}$ to 3.8~$M_{\odot}$ and S/N between $\sim$ 3 and $\sim$ 30. These fields were targeted toward intermediate-to-low column density regions to reveal substructure and investigate whether additional cores could be detected in this regime with deeper observations. All 9 fields overlap with the shallow survey (see Figure~\ref{fig:coverage}); thus, inspecting regions of the cloud that have been observed in both surveys sheds light on how measured core properties, notably for low S/N cores, might change or improve with deeper observations. The deep maps can be seen in greater detail in Figure~\ref{fig:deepmapscontours}, where shallow contours at S/N=2.5 are also shown for reference. We conduct a comparison of cores robustly detected in both surveys to evaluate the flux and size corrections detailed in Section~\ref{sec:coreprop}. To perform this analysis we develop a matching system based on the core footprints, signifying the spatial extent of the core at its 2.5 $\sigma$ detection level boundary. We categorize the cores into three groups: (1) one-to-one isolated matches (2) multiple matches and (3) non-matches, where a core has been detected in the shallow map but has no core coinciding with its location in the deep map. We utilize matched cores to assess the effectiveness of our flux and size correction techniques in both isolated and clustered regions. Additionally, a comparative false detection prediction can be estimated using statistics on shallow non-matched cores.

\subsection{Flux and size correction verification}
Ideally, the deeper observations should trace cores further in their decaying radial surface brightness profile, revealing a higher total flux and larger size.  These measurements should facilitate a direct test of our S/N-based corrections for core flux and area.  We begin by examining the simplistic case of unambiguously matched cores. A core match is classified as unambiguous if the shallow and deep footprints are (1) overlapping (2) the area of the overlapping region is greater than 75\% of the shallow footprint area and (3) the only cores involved in the overlap are a single shallow and single deep core, i.e. there are no extraneous multiple matches in either data set. Throughout the maps there are 14 cores that fit the unambiguous match requirements. These cores have shallow S/N ranging from 3 to 11, with the majority having an improved S/N in the deep data. We apply total flux and half-peak-power (HPP) area corrections to the data, as described in Section~\ref{sec:coreprop}.  

\begin{figure}
\includegraphics[width=.5\textwidth]{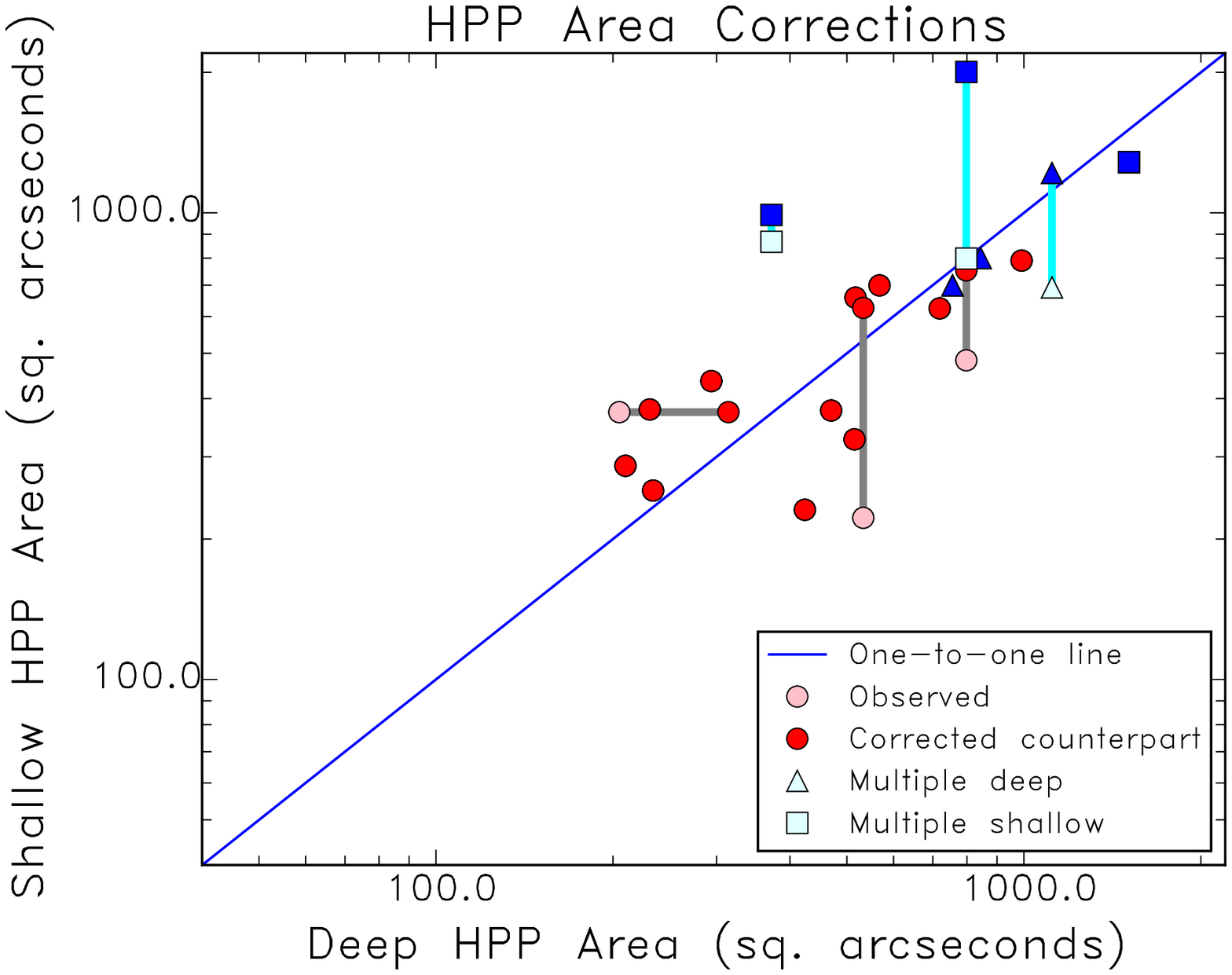}
\includegraphics[width=.5\textwidth]{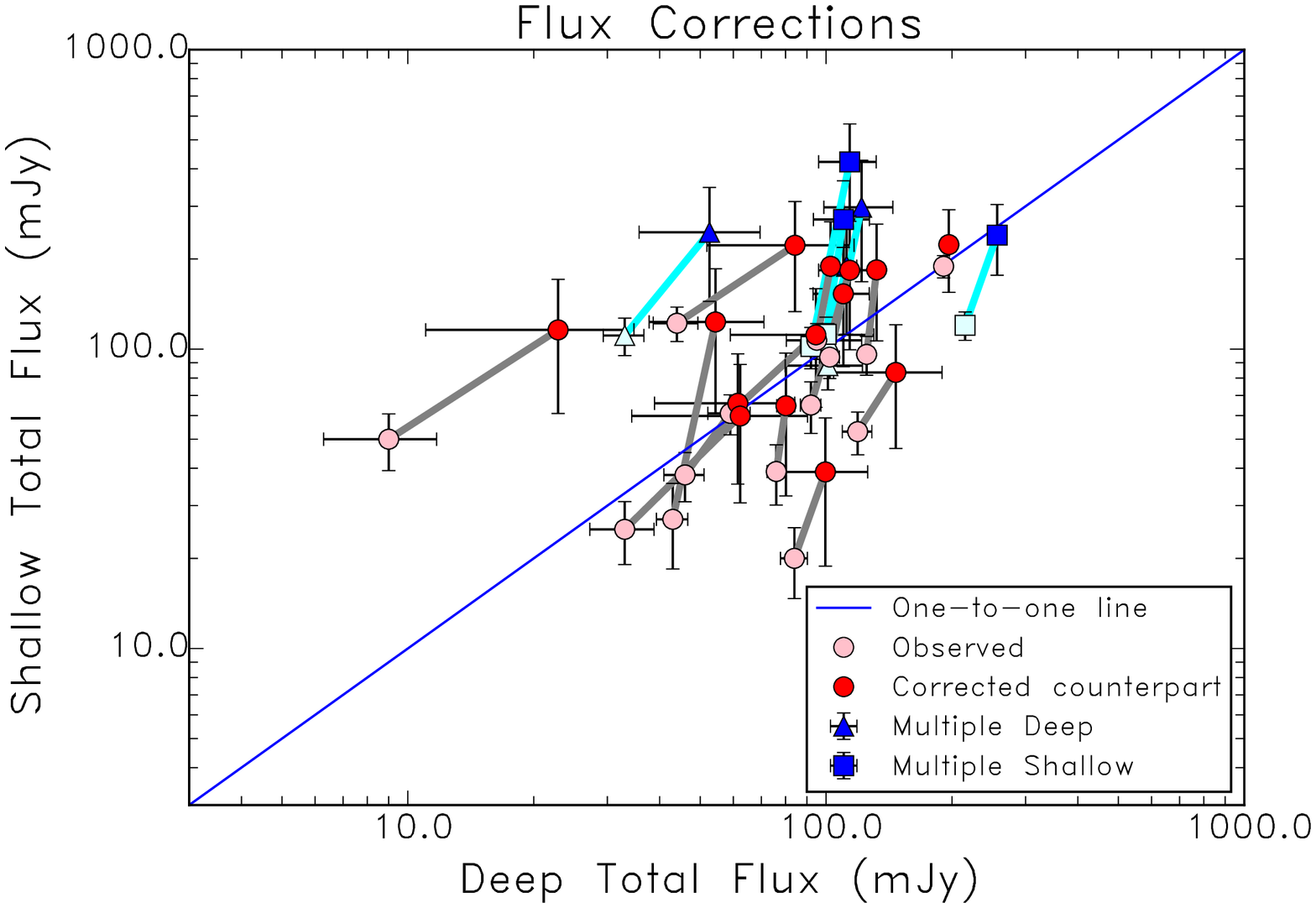}
\caption{Observed and corrected half-peak-power areas and total fluxes and  for the 14 isolated core matches (red) and 6 multiple matches (blue) in both the deep and shallow surveys. For the isolated cores, the light red data points represent observed, uncorrected values from both surveys. These data are connected to their corrected counterparts (dark red) with a grey line, where corrections have been made for both surveys based on methods described in $\S$\ref{sec:corepropFLUX} and $\S$\ref{sec:corepropAREA}. The blue triangles represent summed values of multiple deep cores that matched with a single shallow core while blue squares are summed values for multiple shallow cores that matched with single deep core. Both the uncorrected (light blue) and corrected (dark blue) data are shown, connected with a cyan line. Uncertainties in corrected flux data are a propagated sum of noise and correction errors, while the uncertainties in observed flux data are purely due to noise. The blue line indicates the one-to-one line.}
\label{fig:deepcorr}
\end{figure}

Figure~\ref{fig:deepcorr} shows the comparison between HPP area and total flux in both surveys before and after corrections for the 14 isolated cores in red. The top panel shows measurements of HPP area, where the uncorrected points are connected to their corrected counterparts. The peak-to-total flux ratios of the 79 deep data cores follow a similar S/N distribution as the shallow cores (Figure \ref{fig:peaktotal}). Though the cores exhibit an increase in S/N when observed in the deep survey, the same total flux correction is applied to the deep data based on its S/N. The majority of these points move towards the one-to-one line upon correction.  Observed uncertainties only include measurement noise, while we have adopted a linear uncertainty for the corrected data points of 50\% at S/N=3 to 20\% at S/N= 20 to account for total flux corrections. HPP area (HPPA) corrections are only applied to cores with Peak S/N < 5 and $(\rm HPPA_{corr}- HPPA_{obs})/HPPA_{obs} > 0.5$.  While there are a few outliers in the sample that move parallel to the line, the locality of points near the one-to-one line suggest that the corrections are beneficial to the data for cores that are matched unambiguously.

The next subset of cores have multiple overlapping footprints across the maps. We perform a separate analysis on these matches to determine how the corrections perform in clustered regions and inspect  segmentation noise of cores in low S/N regimes. Core over-segmentation has occurred if multiple peaks in the shallow data have been detected as separate smaller cores when they are in fact a single larger core, as revealed by deeper data. Additionally, cores can be under-segmented if a single shallow core is split into multiple cores in the deep survey.  We find 6 cases of multiple matches in the data. 3 of these involve one shallow core overlapping with multiple deep cores and 3 involve multiple shallow cores corresponding to one deep core. To achieve a fair comparison, we use the total fluxes and sizes of the multiple cores to compare to their 'single' core match properties. As in the analysis of the isolated cores, values in both surveys are corrected as needed and the comparison is shown in Figure~\ref{fig:deepcorr} in blue. 

The multiple match comparison is representative of a regime that is more clustered than the unambiguous matches.  As seen in Figure~\ref{fig:deepcorr}, there is scatter in this comparison around the one-to-one line; however, the spread is roughly symmetric within the sample. While the sum of the corrected values for the multiple-match cores do not always agree across the surveys, we find that correcting these cores introduces a source of noise in the data rather than biasing the results toward one direction. Additionally, within the 6 multiple-match scenarios we find an equal amount of cores (3 each) that have been over-segmented and under-segmented. Although this is a small sample size, the symmetry suggests each segmentation scenario is equally likely throughout the rest of the shallow survey. 

Given the scatter in the multiple-match regime, we investigate whether or not a core's degree of clustering should be considered as a qualifying factor for flux and size corrections. By looking at the distribution of n=2 nearest neighbour distances (NND$_{2}$) i.e. the distance between any core and its closest neighbouring core, we find that the shallow cores involved in multiple matches have NND$_{2}$ < 0.3 pc.  Approximately 62\% of cores in the entire shallow survey have NND$_{2}$ < 0.3 pc, and 7 of the 14 (50\%) unambiguous matches also have a NND$_{2}$ below this value. The unambiguous matches also span a NND$_{2}$ range similar to the entire shallow survey; thus, we are getting a reasonably representative sample of unambiguous matches in all environments.  Since our corrections can perform well in this regime (NND$_{2}$ < 0.3 pc) it is valuable to include them across the range of clustering environments. Finally, within the limitations of our sample comparison, the inclusion of cores that have been over- and under--segmented introduces a symmetric noise in the data without heavily biasing the results.



\begin{figure}
\includegraphics[width=.5\textwidth]{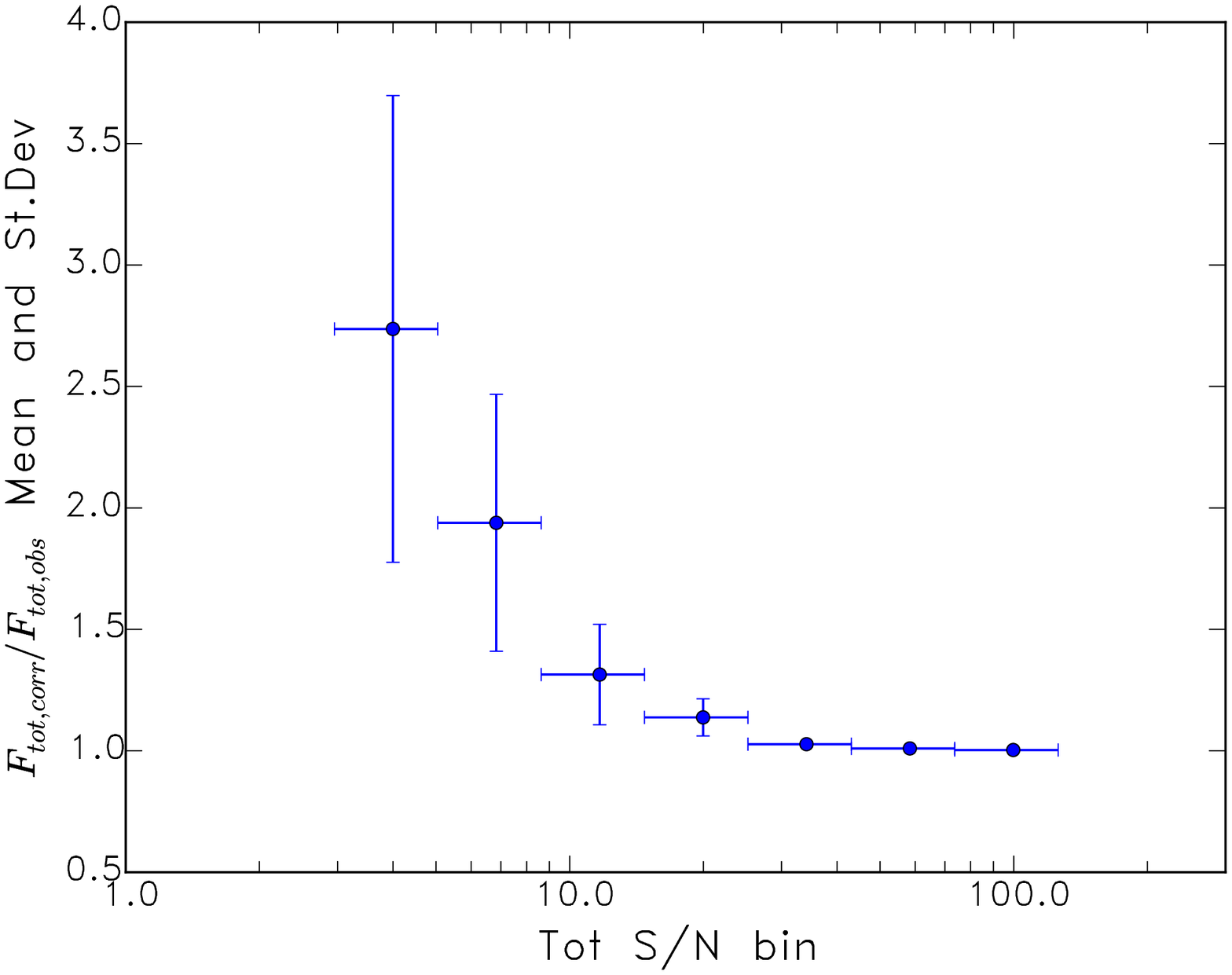}
\caption{The mean total flux correction scale factor, $\frac{F_{tot,corr}}{F_{tot,obs}}$ as a function of total S/N bin. Horizontal error bars represent the S/N bin width while the vertical error bars represent the standard deviation of $\frac{F_{tot,corr}}{F_{tot,obs}}$ in each bin.}
\label{fig:fcorr_vs_sn}
\end{figure}

Based on these tests we conclude that the corrections are still valuable to the analysis and are necessary to prevent claiming falsely small cores at low S/N.  Figure~\ref{fig:fcorr_vs_sn} illustrates the mean and standard deviation total flux correction factor, $\frac{F_{tot,corr}}{F_{tot,obs}}$ as a function of total S/N.  While this factor can be large at low S/N, noise introduced via corrections do not affect the significance of our primary results. The uncertainties in total flux, and therefore mass, faced in crowded regions has a greater impact on the core mass function (CMF) result than the core-gas correlation. To caveat this we have already produced separate CMFs for both isolated and clustered cores (see Figure~\ref{fig:cmfclust}), where the isolated CMF is more robust against core blending uncertainties. The core-gas correlation results are largely unhindered by this effect due to surface mass density smoothing size scales. Despite the possibility of segmentation occurring in crowded regions, once corrected, the total enclosed mass of a given region should remain the same regardless of the number of cores claimed within the area. Additionally, the resultant core-gas correlation is shown to be uniform across varying smoothing size scales in Figure~\ref{fig:smoothing}. Therefore, the consistency of core surface mass density over multiple spatial scales suggests that segmentation does not affect the resulting core-gas correlation.  

\subsection{False detection rate verification}
The final subset of cores are those that were detected in the shallow survey but have no deep counterpart. A total of 10 cores fit this criterion. Despite the small numbers, this method provides a means of testing the false detection estimates derived from our noise realizations. Table~\ref{table:nondetections} details properties of these cores in the shallow map in addition to the corresponding noise at their locations in the deep map, where no core is found. We assume that the deep flux is consistent with zero, thus $\Delta$F $\equiv F_{peak,shal}$ and  $\sigma_{\Delta \rm F}$ is the quadrature sum of the shallow and deep peak noise values at the core location.  We adopt a false core identification threshold such that if $\Delta$F/$\sigma_{\Delta \rm F}$ > 3.0 then the deep data place an upper limit that is inconsistent with the shallow core peak flux, indicating a false detection. 

Of the 10 shallow cores with no deep counterparts, 5 cores have $\Delta$F/$\sigma_{\Delta \rm F}$ > 3.0 and are likely false detections. As described in Section~\ref{sec:coreid}, the false detection estimate within a sample is computed by integrating over the contamination probabilities of the member cores. For the shallow survey, we estimate 21 false candidates of 295 cores and 1 false candidate of 79 cores for the deep survey. To obtain an equivalent estimate for the overlapping data regions, we integrate the contamination probabilities for the 35 shallow cores located in the deep field regions, yielding a predicted estimate of 1.5 false detections in the shallow data. There are more false cores found than this estimate, though the small numbers involved leave the two values statistically consistent.  With that in mind, we note that 4 of the 5 false detection candidates are located in Deep Map 03, a factor of $\sim$30 over-density relative to the other fields combined.  Visually, many shallow mismatches appear close to corresponding deep mismatches in Fig.~\ref{fig:deepcorr}. Also, we note that this field contains a large amount of intermediate scale substructure in the Herschel column density map.  We postulate that these two observations are consistent with conditions that would facilitate relatively strong changes in filtering artifacts because of the different scanning patterns used in each survey (pure raster vs rastered lissajous).   If Deep Map 03 is indeed biased to return anomalously large numbers of false detections through this test, then ignoring them for this exercise gives a false candidate estimate of 1, in excellent agreement with the prediction from the contamination probability integration.

\begin{table*}

\begin{tabular}{|c|c|c|c|c|c|c|c|}

\hline
Shallow Core ID & Shallow Field & Deep Field & Shallow Peak Flux  & Deep Noise  &  $\Delta$F & $\sigma_{\Delta \rm F}$ & $\frac{\Delta F}{\sigma_{\Delta \rm F}}$  \\
                              &             				& 						& (mJy) 						&      (mJy)		& (mJy) &  (mJy) &  \\
\hline
\hline
39 & 03 & 02 & 20.0 & 4.68 & 20.0 & 7.46 & 2.68 \\
40 & 03 & 02 & 21.0 & 4.99 & 21.0 & 7.55 & 2.78 \\
\hline
46 & 04 & 01 & 51.0 & 6.66 & 51.0 & 9.45 & 5.39 \\
48 & 04 & 03 & 51.0 & 3.83 & 51.0 & 7.81 & 6.53 \\
51 &04 & 03 & 37.0 & 1.80 & 37.0 & 7.18 & 5.15 \\
55 &04 & 03 & 29.0 & 2.37 & 29.0 & 7.53 & 3.85 \\
67 & 04 & 03 & 26.0 & 1.27 & 26.0 & 7.34 & 3.54 \\
73 & 04 & 04 & 26.0 & 4.95 & 26.0 & 8.76 & 2.97 \\
\hline
107 & 07 & 09 & 15.0 & 4.00 & 15.0 & 6.12 & 2.45 \\
109 & 07 & 05 & 18.0 & 6.80 & 18.0 & 8.49 & 2.12 \\

\hline

\end{tabular}
\caption{Properties of cores detected in the shallow survey with no coinciding  detection in the deep survey. Deep Noise refers to the mean noise in the deep map at the position of the shallow detection.}
\label{table:nondetections}
\end{table*}


\bsp	
\label{lastpage}
\end{document}